\newif\ifarxiv
\arxivtrue
\ifarxiv\PassOptionsToClass{preprint}{vgtc}\fi

\documentclass{vgtc}                          % final / single-blind
\graphicspath{{figures/}{./}} % where to search for the images

\usepackage{mathptmx}                  % use matching math font

\onlineid{0}

\vgtccategory{Research}

\usepackage{subcaption}

\vgtcinsertpkg

\usepackage[svgnames]{xcolor}   % already loaded by vgtc 'electronic'; same options, so a no-op
\usepackage{booktabs}           % rules in the pattern overview table (tab:patterns)
\usepackage{tikz}               % scene panels in the scenario figure (figures/fig-scenario.tex)

\newif\ifshowcomments
\showcommentstrue

\newcommand{\figplaceholder}[3]{%
  \IfFileExists{figures/#1.pdf}{\includegraphics[width=#2]{figures/#1.pdf}}{%
  \IfFileExists{figures/#1.png}{\includegraphics[width=#2]{figures/#1.png}}{%
  \fbox{\parbox[c][#3][c]{\dimexpr#2-2\fboxsep-2\fboxrule\relax}{%
    \centering\small\sffamily Placeholder: \texttt{figures/#1}}}}}}

\newenvironment{pattern}[2]{\subsubsection{#1}\label{pat:#2}}{\par}

\newcommand{\patternsummary}[1]{\noindent\emph{#1}\par\smallskip}
\newcommand{\patfield}[1]{\par\smallskip\noindent\textbf{#1.}\enspace\ignorespaces}
\newcommand{\patproblem}{\patfield{Problem}}
\newcommand{\patsolution}{\patfield{Solution}}
\newcommand{\patconsequences}{\patfield{Consequences}}
\newcommand{\patknownuses}{\patfield{Known Uses}}

\newlength{\patternwidth}
\newcommand{\scenepanel}[2]{%
  \begin{tikzpicture}
    \IfFileExists{figures/#1.jpg}{%
      \begin{scope}
        \clip[rounded corners=4pt] (0,0) rectangle (\linewidth,0.671\linewidth);
        \node[anchor=south west,inner sep=0] at (0,0)
          {\includegraphics[width=\linewidth]{figures/#1.jpg}};
      \end{scope}}{%
      \node[font=\sffamily\scriptsize] at (0.5\linewidth,0.3355\linewidth)
        {Placeholder: \texttt{figures/#1.jpg}};}
    \draw[rounded corners=4pt,line width=0.5pt,black!60]
      (0,0) rectangle (\linewidth,0.671\linewidth);
    \node[anchor=north west,fill=black!75,text=white,rounded corners=2pt,
          inner sep=2pt,font=\sffamily\scriptsize\bfseries]
      at (3pt,0.671\linewidth-3pt) {#2};
  \end{tikzpicture}}

\newcommand{\pat}[1]{\textsc{#1}}

\ifarxiv
  \preprinttext{Author preprint}             % class default names IEEE VIS
  \hypersetup{pdfsubject={Preprint}}          % class default: VGTC Conference Paper
  \usepackage{tabularx}                       % appendix tables (contents/supp-working-list.tex)
  \newcommand{\suppref}{\cref{sec:appendix:list}}
\else
  \newcommand{\suppref}{the supplemental material}
\fi

\title{Design Patterns for Ubiquitous, Immersive, and Situated Analytics}

\author{Niklas Elmqvist\thanks{e-mail: elm@cs.au.dk}\\ %
        \scriptsize Aarhus University, Denmark %
\and Peter W.\ S.\ Butcher\thanks{e-mail: p.butcher@bangor.ac.uk}\\ %
        \scriptsize Bangor University, United Kingdom %
\and Panagiotis D.\ Ritsos\thanks{e-mail: p.ritsos@bangor.ac.uk}\\ %
        \scriptsize Bangor University, United Kingdom}

\teaser{
  \centering
\begin{minipage}{0.325\linewidth}\scenepanel{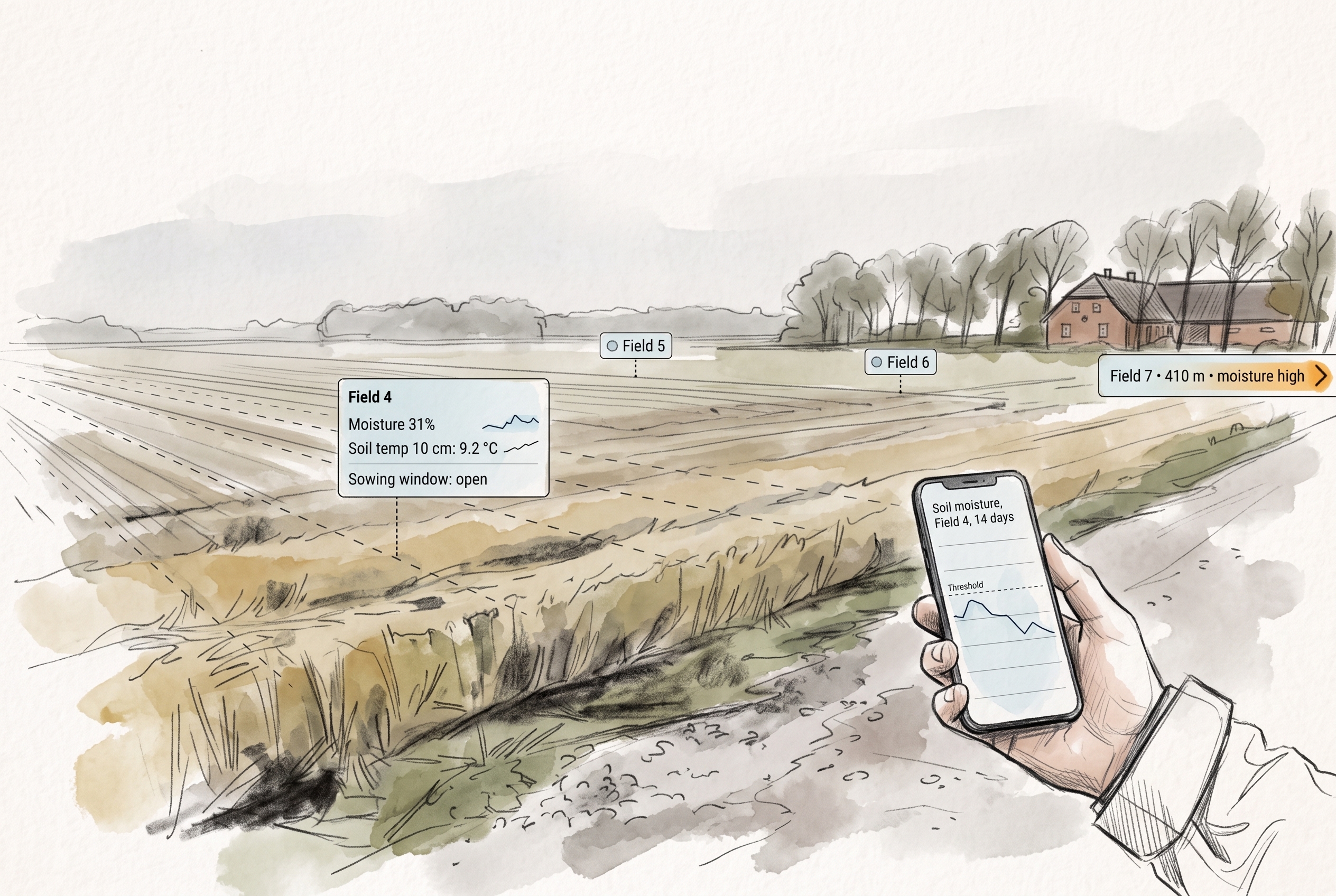}{(a) 07:40}\end{minipage}\hfill
\begin{minipage}{0.325\linewidth}\scenepanel{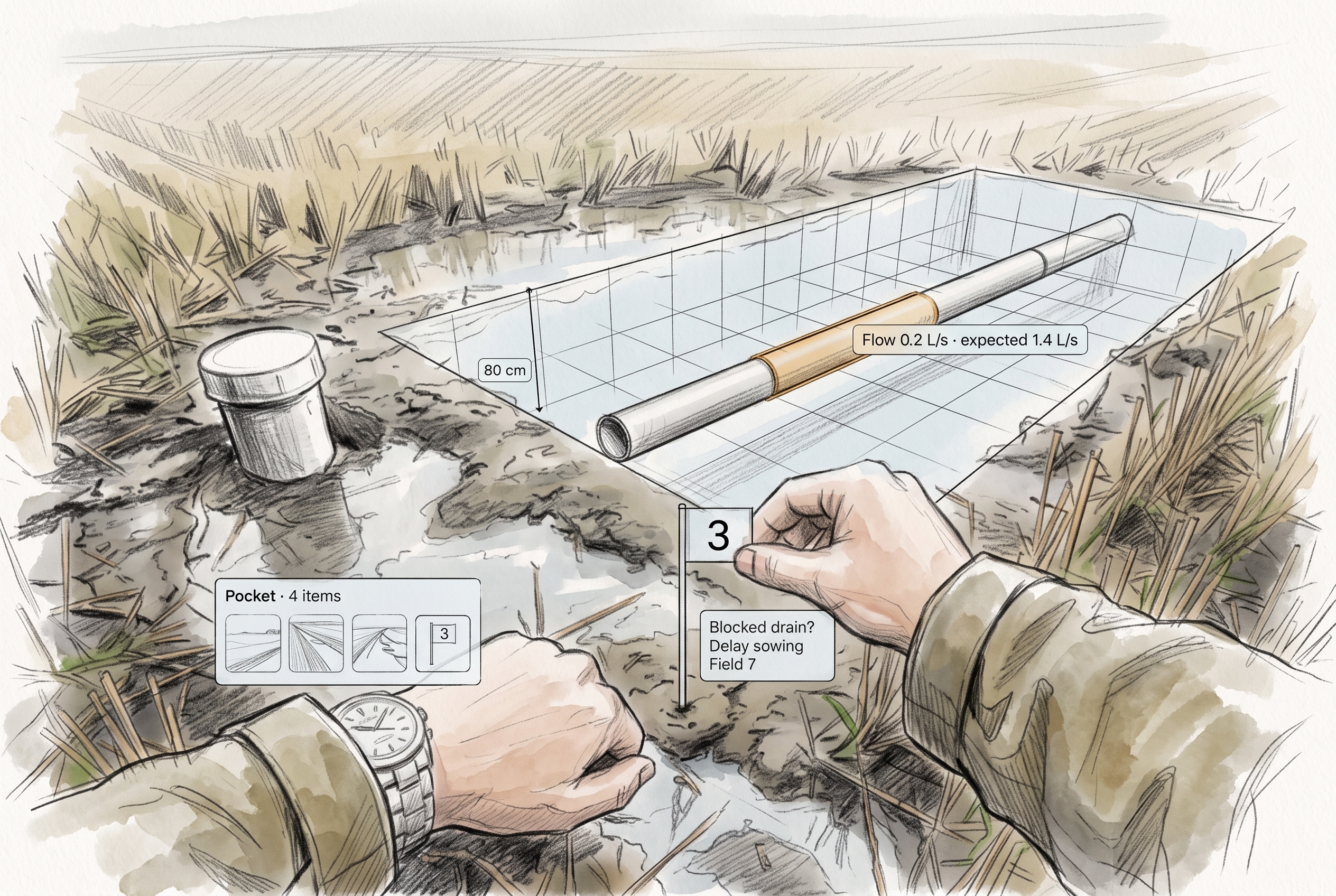}{(b) 08:30}\end{minipage}\hfill
\begin{minipage}{0.325\linewidth}\scenepanel{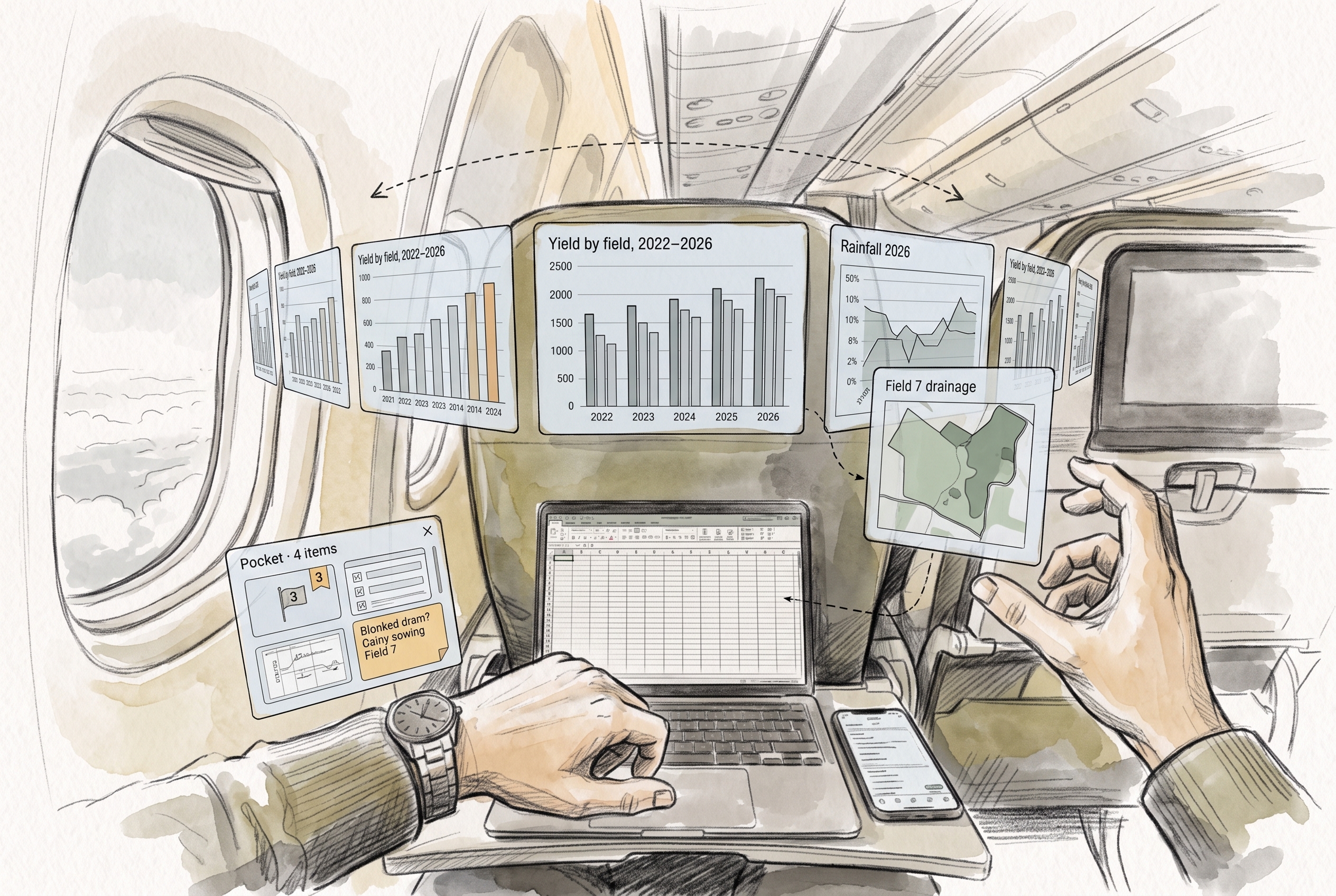}{(c) 14:20}\end{minipage}\par
\vspace{4pt}
\begin{minipage}{0.325\linewidth}\scenepanel{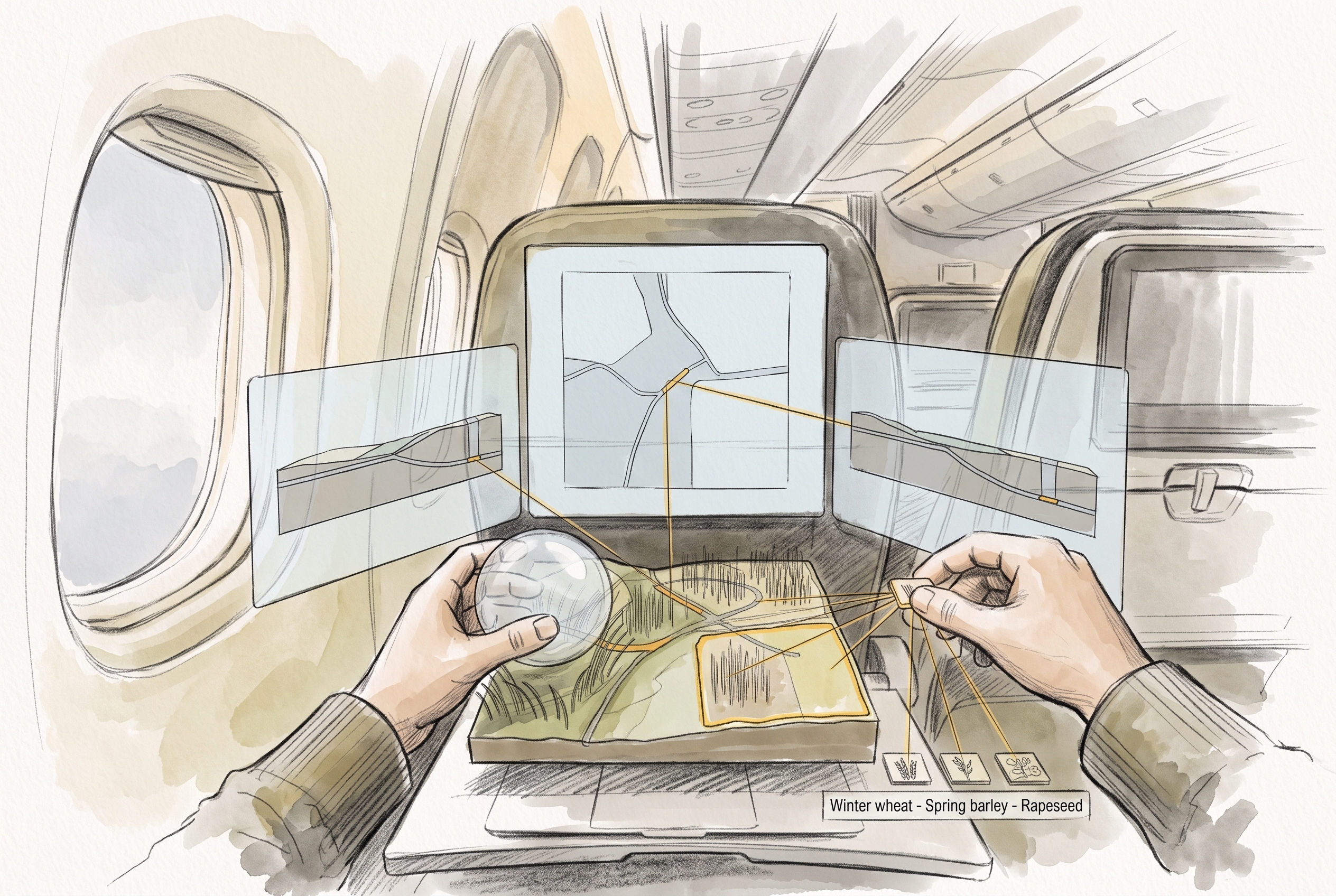}{(d) 14:50}\end{minipage}\hfill
\begin{minipage}{0.325\linewidth}\scenepanel{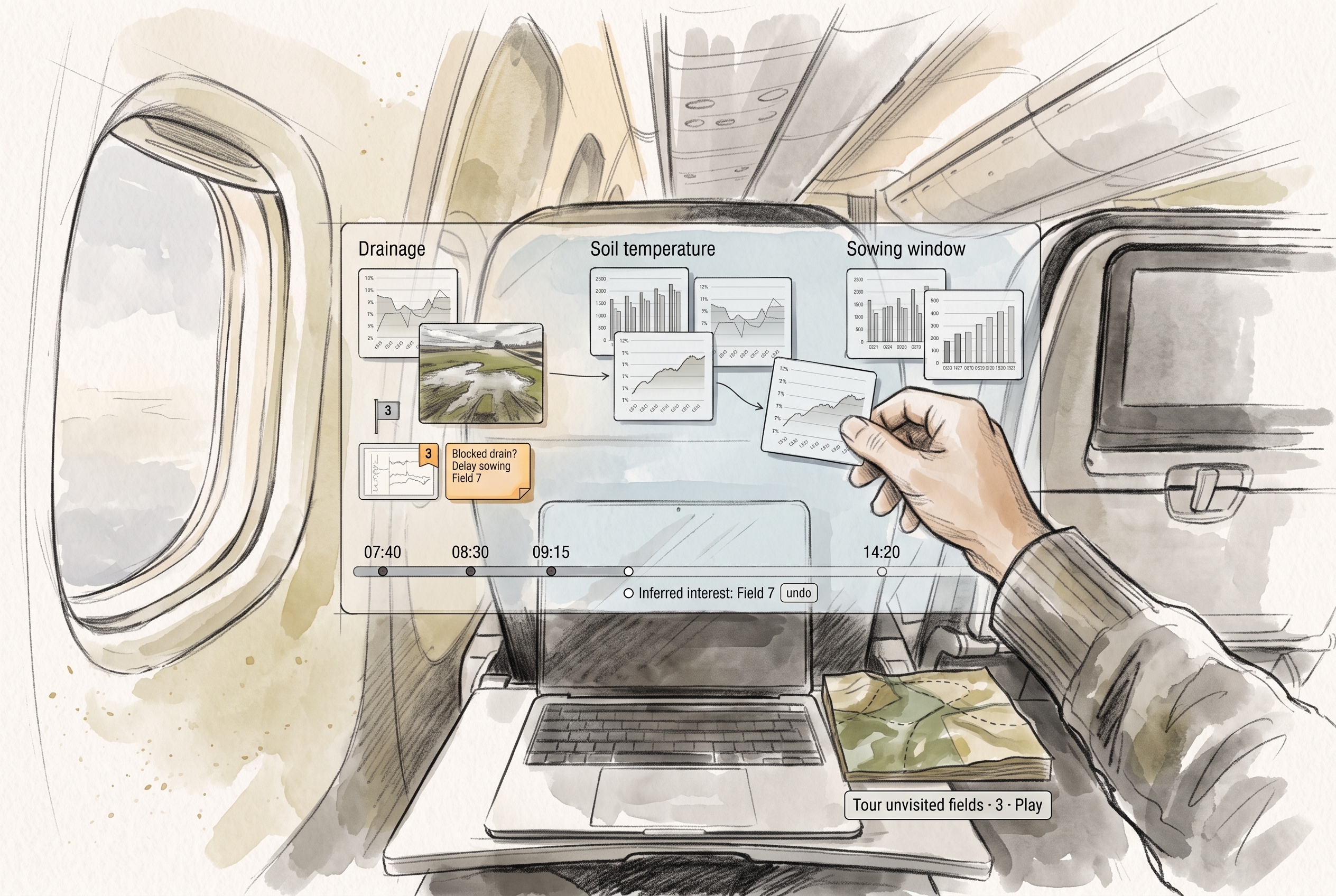}{(e) 15:30}\end{minipage}\hfill
\begin{minipage}{0.325\linewidth}
  \raggedright\scriptsize
  \begin{tabular}{@{}p{1.8em}@{}p{\dimexpr\linewidth-1.8em\relax}@{}}
    (a) & \pat{Bloom}, \pat{Beacon}, \emph{Label}, \emph{Ghost} \\
    (b) & \pat{X-Ray}, \pat{Cairn}, \pat{Pocket} \\
    (c) & \pat{Carousel}, \pat{Pocket}, \pat{Lodestone}, \emph{Panel} \\
    (d) & \pat{Shadow Box}, \pat{Sieve}, \pat{Marionette}, \emph{Proxy} \\
    (e) & \pat{Shoebox}, \pat{Trail}, \pat{Grand Tour}, \emph{Proxy} \\
    \bottomrule
  \end{tabular}\par
  \smallskip
  \textsc{Small caps}: here; \emph{italics}: Lee et al.~\cite{Lee2024}.
\end{minipage}

  \caption{\textbf{Farm analysis in the field.}
  One analyst's day with the patterns: (a, b) on a farm with smart glasses and a phone, and (c--e) on a flight with a mixed-reality headset, a laptop, and a phone.
  (Mockups generated with Google Gemini 3.8 Flash Image.)}
  \label{fig:scenario}
}

\abstract{%
  Visualization designers who move from the desktop to mobile, immersive, and situated settings are greeted by a bewildering array of new platforms, technologies, and terminology. 
Knowledge for designing these applications is scattered across virtual reality, augmented reality, and mobile computing.
We address this gap with a pattern language of 13 design patterns for ubiquitous analytics.
The language extends earlier patterns for situated visualization, which center on visual representation, to interaction at a distance, the analyst's body, and sensemaking. 
We derived the patterns through expert curation, distilling our experience building and studying such systems.
To demonstrate our work, we use all thirteen patterns to design a ubiquitous analytics application from a farm field to an airplane seat.
The resulting vocabulary allows building for these platforms without mastering every discipline the systems span.
}


\begin{document}

%% VGTC conference style: \firstsection must precede \maketitle, so the
%% introduction heading lives here rather than in contents/introduction.tex.
\firstsection{Introduction}
\label{sec:intro}

\maketitle

As data visualization moves from the desktop into mobile, ubiquitous, and pervasive settings, designers need to learn quickly how to build applications that take full advantage of these new platforms.
Unlike the familiar world of mouse and keyboard~\cite{Lee2012}, ubiquitous analytics~\cite{Elmqvist2013, Elmqvist2023} spans an unfamiliar range of displays and input channels, from virtual and augmented reality to gesture and gaze, often under tight mobile power and compute budgets.
The techniques a visualization designer reaches for by habit, such as hover tooltips, brushing and linking across coordinated views, and details on demand, assume a single flat canvas and a cursor; they quietly break when neither is present.
The knowledge for designing these applications exists, but it is scattered.
Some of it lives in immersive analytics, while much more in virtual reality, augmented reality, and mobile computing, where these problems have been studied for decades.
A designer charged with designing a novel visualization tool for these platforms lack access to this body of knowledge.

In this paper, we address this gap with a pattern language of thirteen design patterns for ubiquitous analytics.
Lee et al.~\cite{Lee2024} recently cataloged patterns for situated visualization in augmented reality, centered on how data is represented in the world.
Our language complements theirs by taking a broader view, spanning what the analyst sees, how they reach and act on content at a distance, how they use their own body, and how they record and make sense of what they find.
These four categories---sight, reach, embodiment, and sensemaking---organize the language.
Each pattern names a reusable solution to a problem that recurs across these settings: a \pat{Beacon}, for example, is an in-view marker that points to a referent outside the field of view, while a \pat{Cairn} is a persistent mark the analyst leaves in space to record a finding.
We derived the patterns through expert curation in the tradition of Alexander~\cite{Alexander1977} and the Gang of Four~\cite{Gamma1994}, grounded in systems we have built and studied.

To demonstrate the language, we use it to design a new ubiquitous analytics application that follows an analyst from a farm in the morning to an airplane seat in the afternoon (Figure~\ref{fig:scenario}).
The design leverages all thirteen of our patterns and several of Lee et al.'s.

The contributions of this paper are the following:
(1) a framework that organizes the design space of ubiquitous analytics into four categories---sight, reach, embodiment, and sensemaking---and argues that a pattern language for these settings must span all four;
(2) a language of thirteen design patterns distributed across these categories; and
(3) a design scenario that demonstrates how the language can guide the design of a new ubiquitous analytics application.
We offer the language as an addition to a growing, shared repository of patterns for ubiquitous analytics.

\section{Related Work}
\label{sec:related}

We position our pattern language against three bodies of prior work: research that moved analytics off the desktop, design patterns and the ways their collections were derived, and the design knowledge already codified for immersive and situated settings, where the catalog of Lee et al.~\cite{Lee2024} is closest to ours.

\subsection{Ubiquitous, Immersive, and Situated Analytics}
\label{sec:related:aaa}

Human-computer interaction moved past the desktop long before visualization did.
Weiser~\cite{DBLP:journals/cacm/Weiser93} proposed ubiquitous computing, in which computers leave the desk and spread into the environment.
Post-WIMP research then developed interaction models for interfaces without windows, icons, menus, and a pointer~\cite{vanDam1997, Beaudouin-Lafon2000, Jacob2008}.
Visualization took up the question around 2012 with design considerations beyond mouse and keyboard~\cite{Lee2012} and an interaction model for visualization beyond the desktop~\cite{Jansen2013}.
Elmqvist and Irani~\cite{Elmqvist2013} named the goal ubiquitous analytics, and a year later Roberts et al.~\cite{Roberts2014} called visualization beyond the desktop ``the next big thing.''
By 2022, mobile data visualization had a book of its own~\cite{Lee2022}.

Affordable head-mounted displays then turned the space around the analyst into a canvas for data.
Chandler et al.~\cite{Chandler2015} named this research area immersive analytics in 2015.
A research agenda~\cite{bach16immersiveanalytics}, an edited book~\cite{Marriott2018}, and a set of grand challenges~\cite{Ens2021} followed; several surveys now map the field~\cite{Fonnet2021, Kraus2022, Saffo2024}.
In parallel, augmented reality research tied data to physical places and objects~\cite{ElSayed2015journ}.
SiteLens~\cite{white09sitelens} showed visualizations during urban site visits, situated analytics brought analytical reasoning into augmented reality~\cite{Thomas2018}, and Willett et al.~\cite{willett17embedded} described embedded data representations that sit near or on their physical referents.
Bressa et al.~\cite{Bressa2022} surveyed situatedness in visualization; in earlier work, we surveyed situated analytics~\cite{Shin2024}.
Elmqvist~\cite{Elmqvist2023} gathers these strands under analytics anywhere and everywhere.
Our patterns span all three, including virtual reality, where the data may have no physical referent at all.

\subsection{Design Patterns in Visualization and HCI}
\label{sec:related:patterns}

Design knowledge that must travel between such different settings needs a portable form; the design pattern has provided one for decades.
Alexander and colleagues collected 253 patterns for towns, buildings, and construction, with the theory and method in a companion volume~\cite{Alexander1979, Alexander1977}.
Each pattern names a problem that recurs in a context and the core of its solution.
Beck and Cunningham~\cite{Beck1987} brought pattern languages to object-oriented programming, and Gamma et al.~\cite{Gamma1994} made them mainstream with 23 patterns whose fields (name, problem, solution, consequences, and known uses) our template inherits.
In both fields, experts derived patterns by naming what recurred across many instances.
% Schmidt~\cite{DBLP:journals/cacm/Schmidt95} cautions, from experience with production software, that not every solution should be cast as a pattern; we accordingly triaged our candidates (\cref{sec:method:notpromoted}).

From software, the idea moved into interaction design and visualization.
Borchers~\cite{Borchers2001} proposed a pattern approach to interaction design, while Tidwell~\cite{Tidwell2005} cataloged interface patterns for practitioners.
In visualization, Heer and Agrawala~\cite{Heer2006} described 12 software design patterns for visualization toolkits.
Pousman and Stasko~\cite{Pousman2006} took a data-driven route instead, clustering ambient information systems along their taxonomy into four patterns of design.
Elmqvist and Yi~\cite{DBLP:journals/ivs/ElmqvistY15} carried the idea from design to method with patterns for visualization evaluation.
Bach et al.\ derived patterns for data comics from existing examples and their own practice~\cite{DBLP:conf/chi/BachWFMR18}, and later proposed dashboard design patterns~\cite{DBLP:journals/tvcg/BachFATKFC23}.
Most recently, Dhanoa et al.~\cite{DBLP:journals/cga/DhanoaWLSE25} extracted agent-based design patterns from existing visualization systems through expert analysis.

\subsection{Design Patterns for Situated Visualization}
\label{sec:related:lee}

As immersive and situated systems multiplied, researchers began to codify what works in them.
Design spaces address 2D information in 3D mixed reality~\cite{ens14etherealplanes}, spatial user interfaces for in-situ visual analytics~\cite{Ens2017}, and immersive urban analytics~\cite{chen17urbananalytics}.
Taxonomies and surveys cover 3D occlusion management~\cite{Elmqvist2008}, interaction for immersive analytics~\cite{DBLP:series/lncs/BuschelCDDDG0KNS18}, spatial interfaces for 3D visualization~\cite{DBLP:journals/cgf/BesanconYKY021}, hybrid cross-device user interfaces~\cite{hubenschmid2025hybriduserinterfacespast}, and the design of situated interaction in public spaces~\cite{DBLP:conf/chi/Zhu2024misitu}.
A design space maps the dimensions along which designs vary, while a pattern names one reusable solution to a recurring problem; the two complement each other.

Lee et al.~\cite{Lee2024} took the step from a single architectural pattern to a full catalog.
Their ten patterns for situated visualization in augmented reality, among them Proxy and Magic Lens, center on how data is represented relative to its physical referent.
The authors screened more than 2,000 papers down to 457; the senior author free-coded this corpus, and the other two authors assigned the papers to the resulting patterns.
This corpus-grounded expert coding gave designers of augmented reality visualizations their first shared vocabulary.
Our catalog builds on theirs: we used their patterns as a reference during triage (\cref{sec:method:curation}), and several of our entries state their relation to one of them, as \pat{Marionette}~(\cref{pat:marionette}) does for Proxy and \pat{Sieve}~(\cref{pat:sieve}) does for Magic Lens~\cite{Lee2024}.

Our language extends this shared vocabulary by widening the scope from visual representation to four categories (sight, reach, embodiment, and sensemaking) and covers mobile, cross-device, and fully virtual settings, where the data may have no physical referent.
The two languages compose: the design scenario in \cref{sec:scenario} combines patterns from both.
Together, the two catalogs form a shared pattern language for ubiquitous/immersive/situated analytics.

\section{Method}
\label{sec:method}

Our pattern language was designed to complement existing pattern catalogs for ubiquitous, immersive, and situated analytics applications.
Rather than focusing on visual representations alone, we cast a broad net to also discuss interaction, sensemaking, and embodiment.
Finally, in the grand tradition of design patterns, we derived ours through expert curation and by drawing on long experience.

\subsection{Goal and Scope}
\label{sec:method:scope}

The goal of our analysis was to include interaction and analytics across mobile, cross-device, and immersive settings.
Where one of our patterns touches existing patterns, we describe the difference.

\subsection{Curation Process}
\label{sec:method:curation}

We derived the patterns through expert curation and a targeted literature review.
The three authors drew up a candidate list from their own research and from prior systems, then refined it over two weeks through synchronous meetings and asynchronous discussion.
During triage, we applied three tests to each candidate: whether it overlapped with one of the patterns of Lee et al.~\cite{Lee2024}, whether it described the same design move as another candidate and could merge with it, and whether a narrow technique could generalize into a broader pattern.
The merge test, for example, combined gaze tracking and distance-based visual representations into \pat{Bloom}, and the overlap test split World in Miniature~\cite{stoakley95virtual} between our \pat{Marionette} and the Proxy pattern of Lee et al.
A candidate became a pattern only if it had at least two known uses.

In the tradition of pattern languages written by practitioners~\cite{Alexander1977, Gamma1994}, the curation rests on our experience building and studying these systems.
Elmqvist built occlusion techniques among the known uses of \pat{Beacon}, \pat{Sieve}, and \pat{X-Ray}~\cite{DBLP:conf/interact/JavedKGE11, Elmqvist2005, Elmqvist2007d} and the camera tours behind \pat{Grand Tour}~\cite{DBLP:conf/chi/ElmqvistTT08, DBLP:conf/vrst/ElmqvistTT07}.
He also coauthored the cross-device, proxemic, and storytelling systems cited under \pat{Pocket}, \pat{Bloom}, and \pat{Trail}~\cite{Horak2018dng, Badam2016b, DBLP:journals/ivs/ParkSZDRE22}.
Elmqvist, Butcher, and Ritsos together built the ubiquitous analytics systems cited under \pat{Shadow Box}, \pat{Grand Tour}, \pat{Lodestone}, and \pat{Bloom}~\cite{BatchCGF2023, Borowski2025spatialstrates, Borowski2025dashspace, DBLP:journals/tvcg/SrinivasanEBRE25}.

The working list contained 39 entries: the ten patterns of Lee et al.~\cite{Lee2024}, which we kept as a baseline, and 29 candidates of our own in eight groupings (see \suppref).
Triage reduced our 29 candidates to thirteen patterns in four categories.
\Cref{sec:method:notpromoted} accounts for the candidates we dropped or absorbed.

\subsection{Pattern Format}
\label{sec:method:template}

Each pattern follows the tradition of Alexander~\cite{Alexander1977} and of Gamma et al.~\cite{Gamma1994}.
An entry gives the pattern's Name, Problem, Solution, Consequences, and Known Uses, together with a brief summary.

\subsection{Categories}
\label{sec:method:categories}

We group the patterns into four categories.
\emph{Sight} patterns let the analyst see what is hidden by distance, occlusion, or depth.
\emph{Reach} patterns close the distance between the analyst and the content: the analyst acts on it from afar, goes to it, or brings it closer.
In the \emph{embodiment} patterns, the analyst's body becomes an anchor, a container, or an input, and the \emph{sensemaking} patterns help the analyst record, organize, and retrace findings.

\subsection{Pattern Candidates}
\label{sec:method:notpromoted}

Several ideas for patterns were discussed but not incorporated into our final corpus.
Our working list contained 39 entries, the ten patterns of Lee et al.\ and 29 candidates of our own; \suppref{} lists and briefly describes all of them.

\section{Design Patterns}
\label{sec:patterns}

We organize the thirteen patterns into four categories, summarized in \cref{tab:patterns}.
Each entry follows the format described in \cref{sec:method:template}.

\begin{table*}[tb]
  \caption{\textbf{Pattern language.} Thirteen design patterns for anytime anywhere analytics grouped by category.}
  \label{tab:patterns}
  \centering
  \small
  \begin{tabular}{@{}llp{0.66\textwidth}@{}}
    \toprule
    \textbf{Category} & \textbf{Pattern} & \textbf{Summary} \\
    \midrule
    Sight       & \pat{Beacon}     & An in-view marker that points to a referent outside the field of view or hidden behind something else. \\
                & \pat{Sieve}      & A movable volume that removes, pushes aside, or fades whatever it passes through, so the analyst can see past clutter. \\
                & \pat{X-Ray}      & Occluders turn transparent when the data marks something behind them as important. \\
                & \pat{Shadow Box} & The 3D scene unfolds into a set of flat orthographic views arranged around it for detailed inspection. \\
    \midrule
    Reach       & \pat{Marionette} & A nearby stand-in that, when manipulated, acts on distant or grouped objects. \\
                & \pat{Grand Tour} & The system moves the camera along a path that shows everything worth seeing, leaving the scene untouched. \\
                & \pat{Lodestone}  & The analyst attracts content to the body or repels it away, moving the scene instead of the viewer. \\
    \midrule
    Embodiment  & \pat{Carousel}   & Items rotate past a fixed viewing position, so the analyst browses a collection without walking. \\
                & \pat{Pocket}     & A personal container carried on the body that holds views and data across places and devices. \\
                & \pat{Bloom}      & Content reveals more detail as the analyst approaches or attends to it, and closes again as attention moves on. \\
    \midrule
    Sensemaking & \pat{Cairn}      & The analyst leaves a persistent mark at a point, region, or plane to record a finding in place. \\
                & \pat{Shoebox}    & A spatial workspace where the analyst collects findings and arranges them into a structure. \\
                & \pat{Trail}      & A visible, revisitable record of the analysis path, covering both the analyst's actions and the system's inferences. \\
    \bottomrule
  \end{tabular}
\end{table*}

\subsection{Sight}
\label{sec:patterns:sight}

% =====================================================================
% Category figure: Sight, subfigures (a) to (c)
% Subfigure (d) continues in figures/fig-sight-continued.tex.
% Input from contents/patterns.tex at the start of the Sight subsection.
% =====================================================================

\begin{figure*}[t]
  \centering
  \begin{subfigure}[t]{\patternwidth}
    \includegraphics[width=\linewidth]{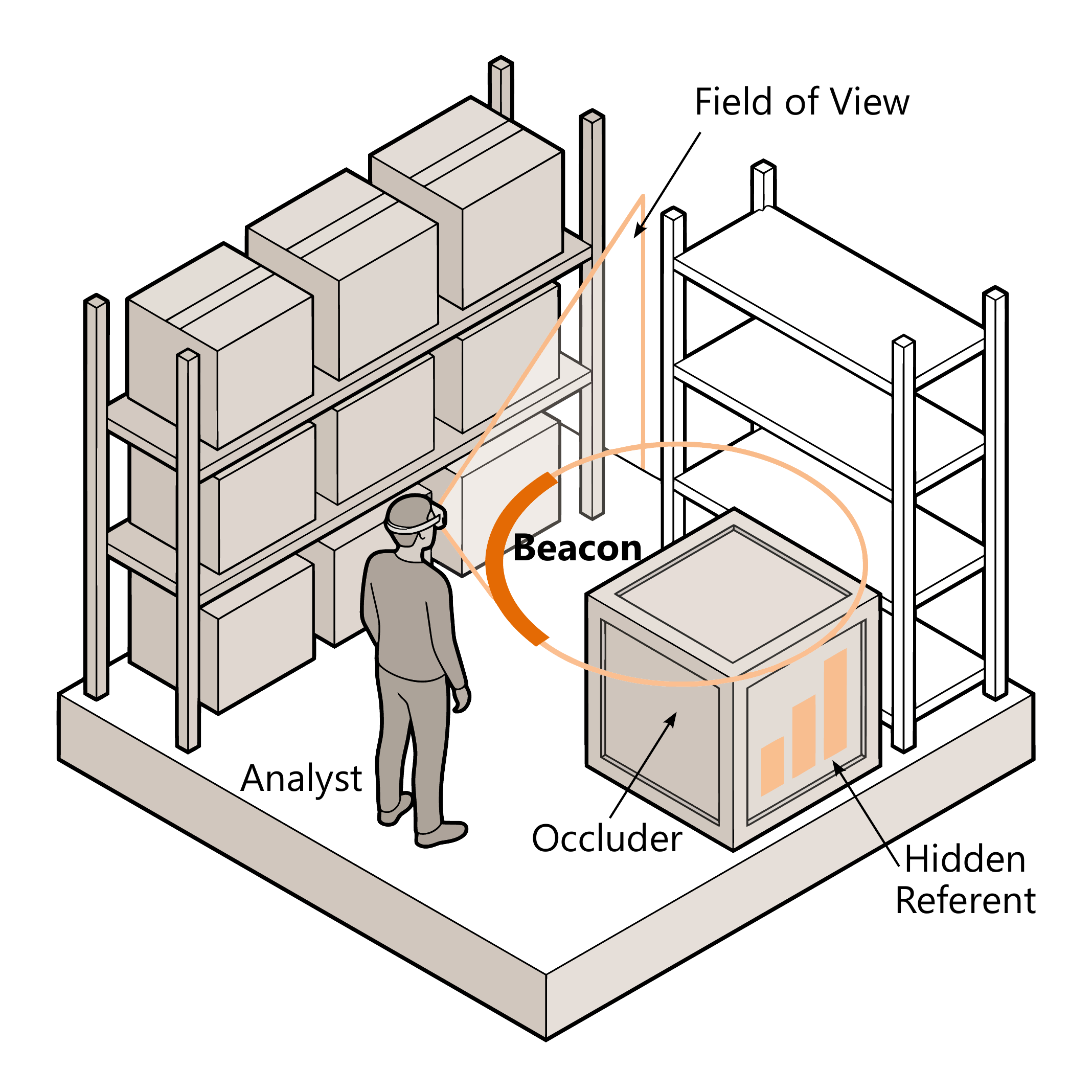}
    \caption{\pat{Beacon}.
    An arc-shaped marker at the edge of the analyst's field of view points toward a part of a crate hidden from view.}
    \label{fig:pat:beacon}
  \end{subfigure}\hfill
  \begin{subfigure}[t]{\patternwidth}
    \includegraphics[width=\linewidth]{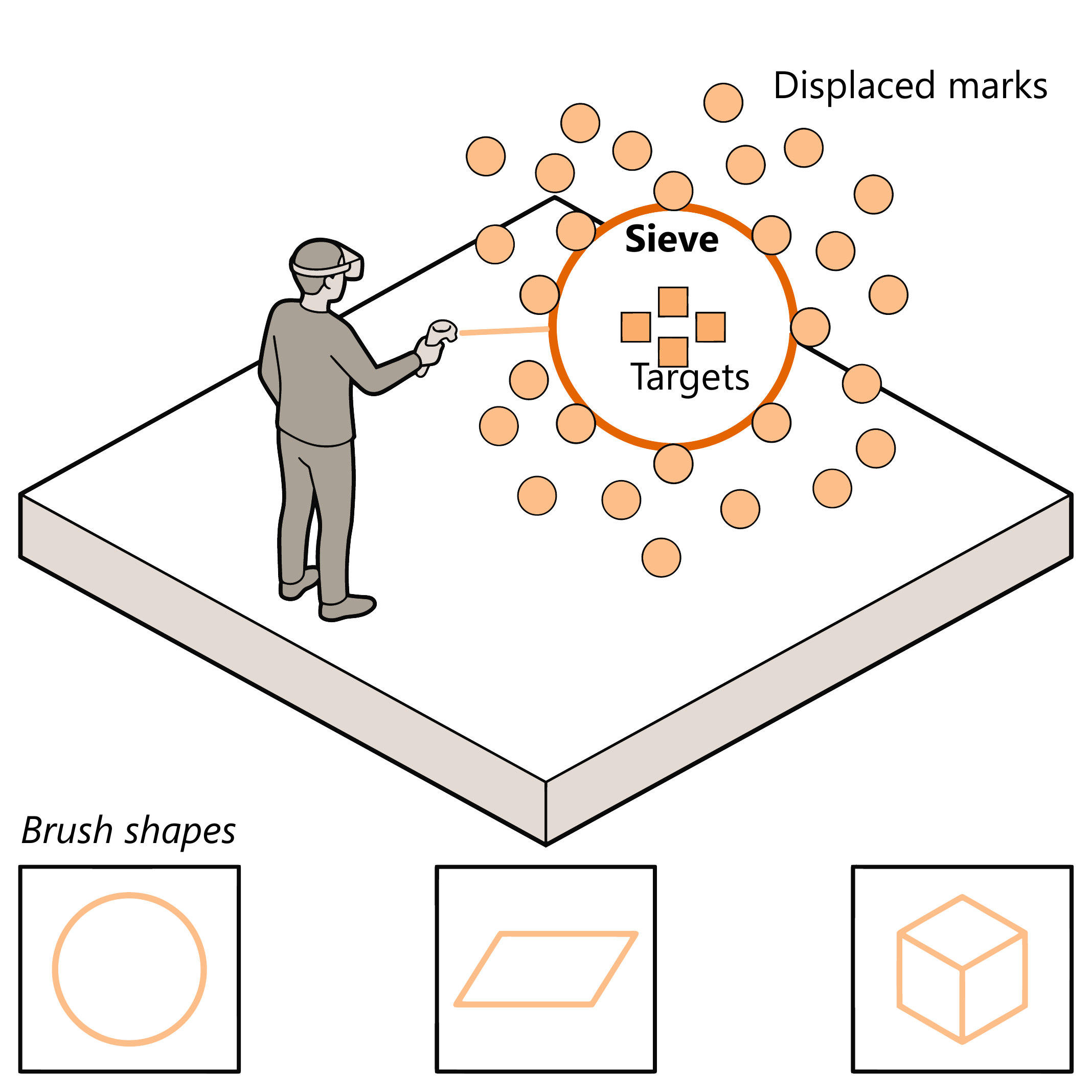}
    \caption{\pat{Sieve}.
    A spherical brush pushes the data points it passes through onto its surface, opening a cavity around a target cluster.
    Insets show alternative brush shapes.}
    \label{fig:pat:sieve}
  \end{subfigure}\hfill
  \begin{subfigure}[t]{\patternwidth}
    \includegraphics[width=\linewidth]{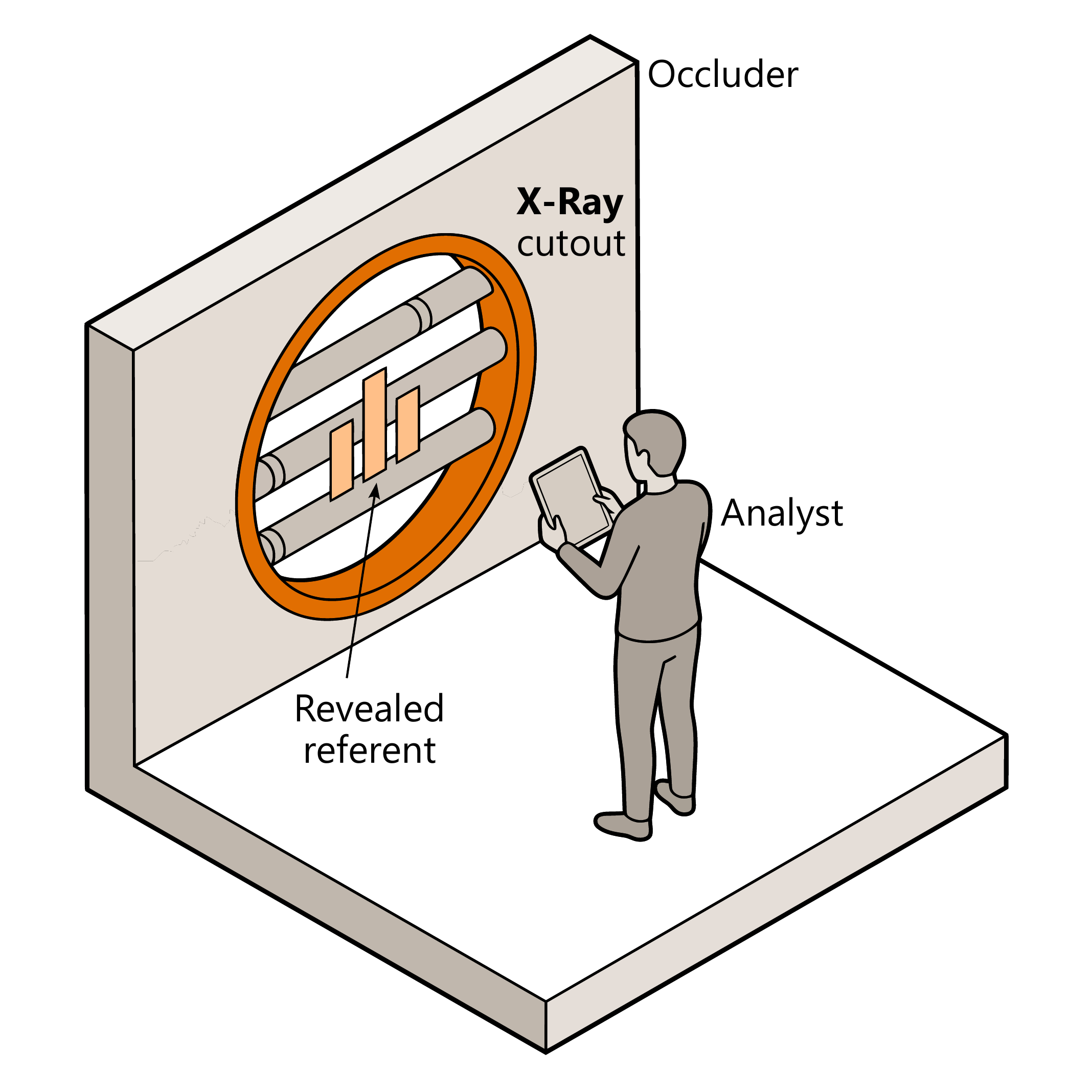}
    \caption{\pat{X-Ray}.
    A circular cutaway in a wall reveals the pipes behind it.
    Only the segment that matches a data condition carries data glyphs; the rest of the wall stays opaque.}
    \label{fig:pat:xray}
  \end{subfigure}
  \caption{\textbf{Sight patterns.}
    Each pattern shows the analyst something that distance, occlusion, or depth hides: a referent out of view~(a), a target inside clutter~(b), or content behind an occluder~(c).
    \pat{Shadow Box} completes the category in \cref{fig:pat:shadow-box}.
    In all pattern figures, black outlines and gray fills mark physical things, pale orange marks virtual content, and saturated orange marks the element that defines the pattern.}
  \label{fig:patterns-sight}
\end{figure*}

Sight patterns let the analyst see what distance, occlusion, or depth hides.
\pat{Beacon} tells the analyst that a hidden referent exists and where to find it.
\pat{Sieve} and \pat{X-Ray} both clear occluders out of the way, the first by position and the second by what the data says lies behind them.
\pat{Shadow Box} gives up depth for a set of 2D views.

% =====================================================================
% Pattern: Beacon
% Category: Sight
% Source: notes/patterns.md
% =====================================================================

\begin{pattern}{Beacon}{beacon}

\patternsummary{An in-view marker that points to a referent outside the field of view or hidden behind something else.}

\patproblem
Headsets tend to have a narrow field of view, and so do hand-held Augmented Reality realized through a mobile device that the analyst holds up to a scene.
Dense physical environments add many objects that can hide a referent.
The referent of interest then lies outside the field of view or behind another object, so the analyst does not know that it exists or where to find it.

\patsolution
Place a small marker inside the view, either at the edge of the view or on the occluder.
The marker encodes the direction of the hidden referent and, optionally, its distance and identity.

\patconsequences
Beacons compete for edge space as their number grows, so a large set of beacons needs aggregation.
Each beacon also adds visual complexity, which the designer must weigh against the cost of missing a referent.

\patknownuses
Halo~\cite{Baudisch2003} and Wedge~\cite{DBLP:conf/chi/GustafsonBGI08} point to off-screen objects from the edge of a 2D display.
Javed et al.\ manage occlusion on tabletop displays~\cite{DBLP:conf/interact/JavedKGE11} with degrees of awareness and access.
EyeSee360 provides off-screen awareness on head-worn displays~\cite{DBLP:conf/sui/GruenefeldEAHB17}.

\end{pattern}

% =====================================================================
% Pattern: Sieve
% Category: Sight
% Source: notes/patterns.md
% =====================================================================

\begin{pattern}{Sieve}{sieve}

\patternsummary{A movable volume that removes, pushes aside, or fades whatever it passes through, so the analyst can see past clutter.}

\patproblem
Volumetric data, dense point clouds, and cluttered physical or virtual scenes pack many objects into a 3D space.
Many of these objects occlude the ones the analyst wants to inspect.
The occluders share a region of space but no common attribute, so filtering by attribute does not isolate them.

\patsolution
Give the analyst an operator that moves through the space and acts on whatever falls inside it.
The pattern has four slots: the brush geometry (sphere, plane, box, etc), its placement, what it selects, and its effect (displace, fade, or remove).

\patconsequences
Displacement preserves context but distorts spatial relations, removal loses context, and fading keeps both at the price of visual noise.
A capacity limit, or view budget, can determine the selection slot.
The pattern clears the region it covers; an operator that restyles what it contains belongs to the Magic Lens pattern~\cite{Lee2024, Bier1993}.

\patknownuses
McGuffin et al.\ deform volumetric data so the analyst can browse its interior~\cite{DBLP:conf/visualization/McGuffinTB03}.
BalloonProbe pushes occluding objects aside with an interactive sphere or wedge~\cite{Elmqvist2005}.
Cutting planes in volume visualization, such as for MRI imaging, remove everything on one side of a plane.
3D Magic Lenses~\cite{DBLP:conf/uist/ViegaCWP96} count as Sieves when they filter what they contain.

\end{pattern}

% =====================================================================
% Pattern: X-Ray
% Category: Sight
% Source: notes/patterns.md
% =====================================================================

\begin{pattern}{X-Ray}{xray}

\patternsummary{Occluders turn transparent when the data marks something behind them as important.}

\patproblem
Situated AR overlays physical structures, whereas immersive scenes often use meaningful spatial layouts that must remain intact.
In both settings, important content sits behind objects that cannot or should not be moved, such as walls, machinery, or the body.
The analyst often needs to see that content despite occluders.

\patsolution
Render occluders transparent, or cut them away, when the content behind them satisfies a data predicate.
The predicate can be a value threshold, a filter match, a selection, or a query result.

\patconsequences
Depth perception suffers when occluders vanish, so the pattern needs depth cues such as edges or partial opacity.
X-Ray takes its predicate from the data, whereas a \pat{Sieve}~(\cref{pat:sieve}) acts on whatever lies inside a movable operator.

\patknownuses
Bane and H\"{o}llerer provide virtual x-ray vision in mobile AR~\cite{DBLP:conf/ismar/BaneH04}, and Coffin and H\"{o}llerer render interactive perspective cut-away views~\cite{DBLP:conf/3dui/CoffinH06}.
Avery et al.\ study how to preserve depth perception in AR x-ray vision~\cite{DBLP:conf/vr/AveryST09}.
Elmqvist et al.\ apply dynamic transparency to manage occlusion in 3D~\cite{Elmqvist2007d}.

\end{pattern}

% =====================================================================
% Category figure: Sight (continued), subfigure (d)
% Continues figures/fig-sight.tex under the same figure number.
% Input from contents/patterns.tex at the start of the fourth Sight pattern.
% =====================================================================

% TODO(NE): Shadow Box sits here because it is the last Sight pattern in document order; swap it with a subfigure of fig-sight.tex to change which pattern stands alone.
\begin{figure}[t]
  \ContinuedFloat
  \centering
  \begin{subfigure}[t]{\patternwidth}
    \includegraphics[width=\linewidth]{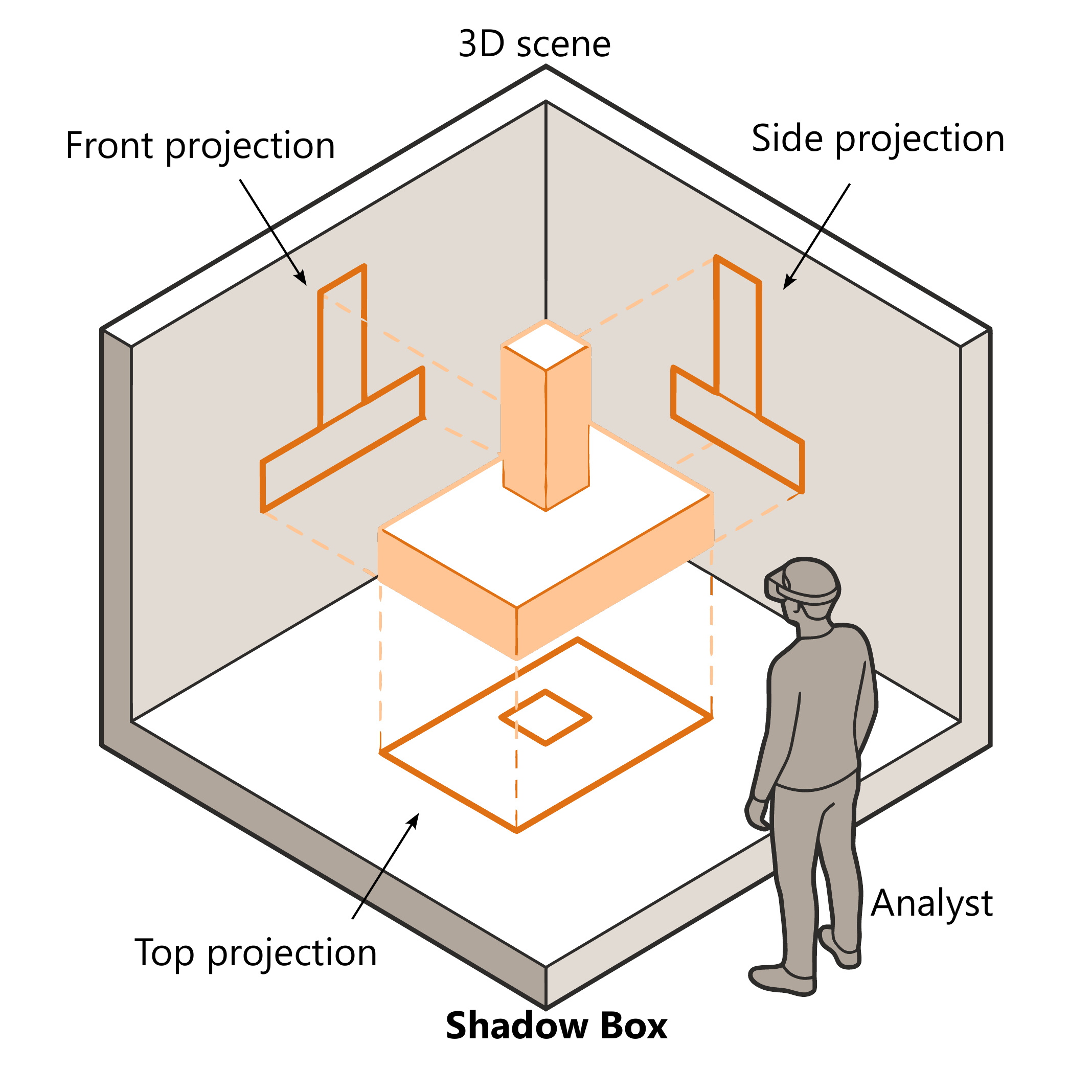}
    \caption{\pat{Shadow Box}.
    A floating 3D model.
    The floor, back wall, and side wall show its plan, front elevation, and side elevation, with projection lines back to the model.}
    \label{fig:pat:shadow-box}
  \end{subfigure}
  \caption[]{\textbf{Sight patterns} (continued).}
\end{figure}

% =====================================================================
% Pattern: Shadow Box
% Category: Sight
% Source: notes/patterns.md
% =====================================================================
\begin{pattern}{Shadow Box}{shadow-box}

\patternsummary{The 3D scene unfolds into a set of flat orthographic views arranged around it for detailed inspection.}

\patproblem
Immersive and desktop 3D scenes often have layered or vertical structure, such as buildings, volumes, and stacked views.
Depth and perspective hide much of the detail in these scenes.
A 2D overview map flattens the scene and loses its vertical dimension.

\patsolution
Unfold the scene into several orthographic projections arranged on the faces of a box around it.
Let the analyst animate between the 3D view and the unfolded one.

\patconsequences
The analyst must link positions across projections, which brushing and leader lines support.
The Proxy pattern of Lee et al.~\cite{Lee2024} covers a single reduced 2D map, and Shadow Box differs from it by unfolding the scene onto multiple faces.

\patknownuses
Herndon et al.'s Interactive Shadows project 3D objects onto the walls and floor, where the user can manipulate the projections directly~\cite{DBLP:conf/uist/HerndonZRCSD92}.
ExoVis places 2D slices and projections around a 3D volume~\cite{DBLP:conf/graphicsinterface/ToryS03}, and Tory et al.\ combine 2D and 3D displays~\cite{DBLP:journals/tvcg/ToryKAM06}.
Engineering drawing follows the same principle in multiview orthographic projection.
In immersive analytics, Batch et al.\ build a shadow box~\cite{BatchCGF2023}, and SpatialStrates~\cite{Borowski2025spatialstrates} and DashSpace~\cite{Borowski2025dashspace} arrange views on volumes.

\end{pattern}

\subsection{Reach}
\label{sec:patterns:reach}

% =====================================================================
% Category figure: Reach, subfigures (a) to (c)
% Input from contents/patterns.tex at the start of the Reach subsection.
% =====================================================================

\begin{figure*}[t]
  \centering
  \begin{subfigure}[t]{\patternwidth}
    \includegraphics[width=\linewidth]{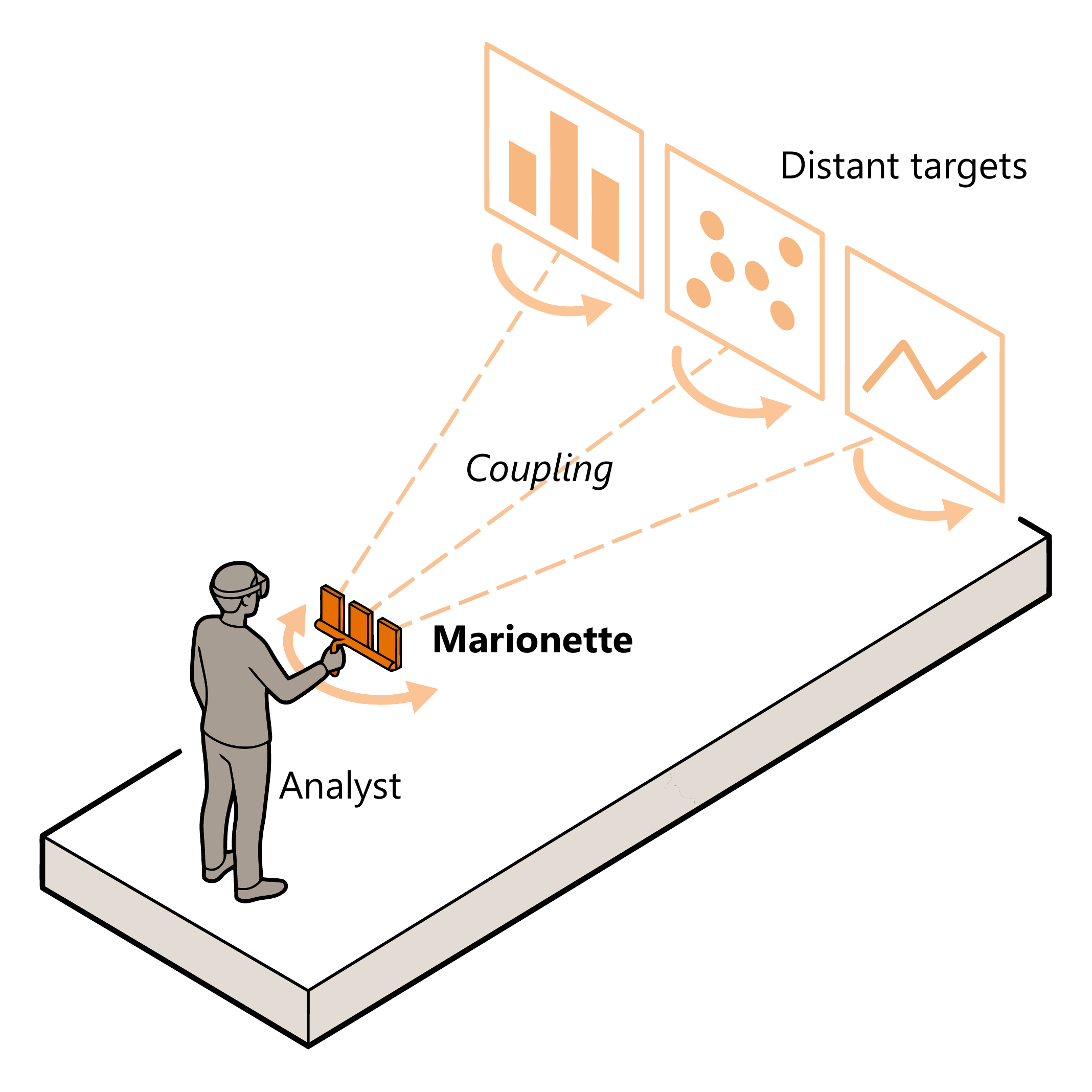}
    \caption{\pat{Marionette}.
    The analyst rotates a small handle, and three distant chart panels coupled to it rotate in unison.}
    \label{fig:pat:marionette}
  \end{subfigure}\hfill
  \begin{subfigure}[t]{\patternwidth}
    \includegraphics[width=\linewidth]{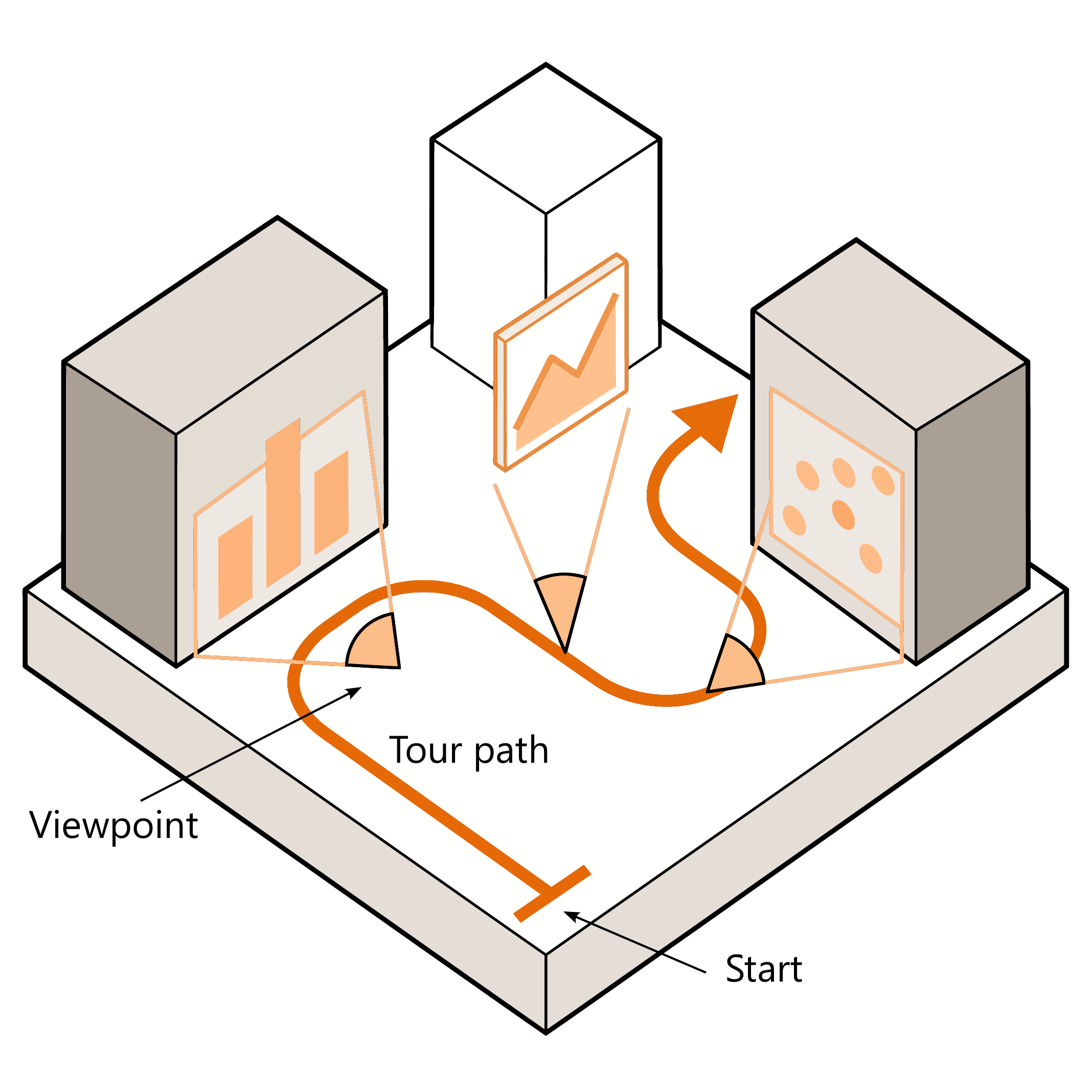}
    \caption{\pat{Grand Tour}.
    A camera path winds through a scene.
    The viewpoints along the path together cover every data glyph.}
    \label{fig:pat:grand-tour}
  \end{subfigure}\hfill
  \begin{subfigure}[t]{\patternwidth}
    \includegraphics[width=\linewidth]{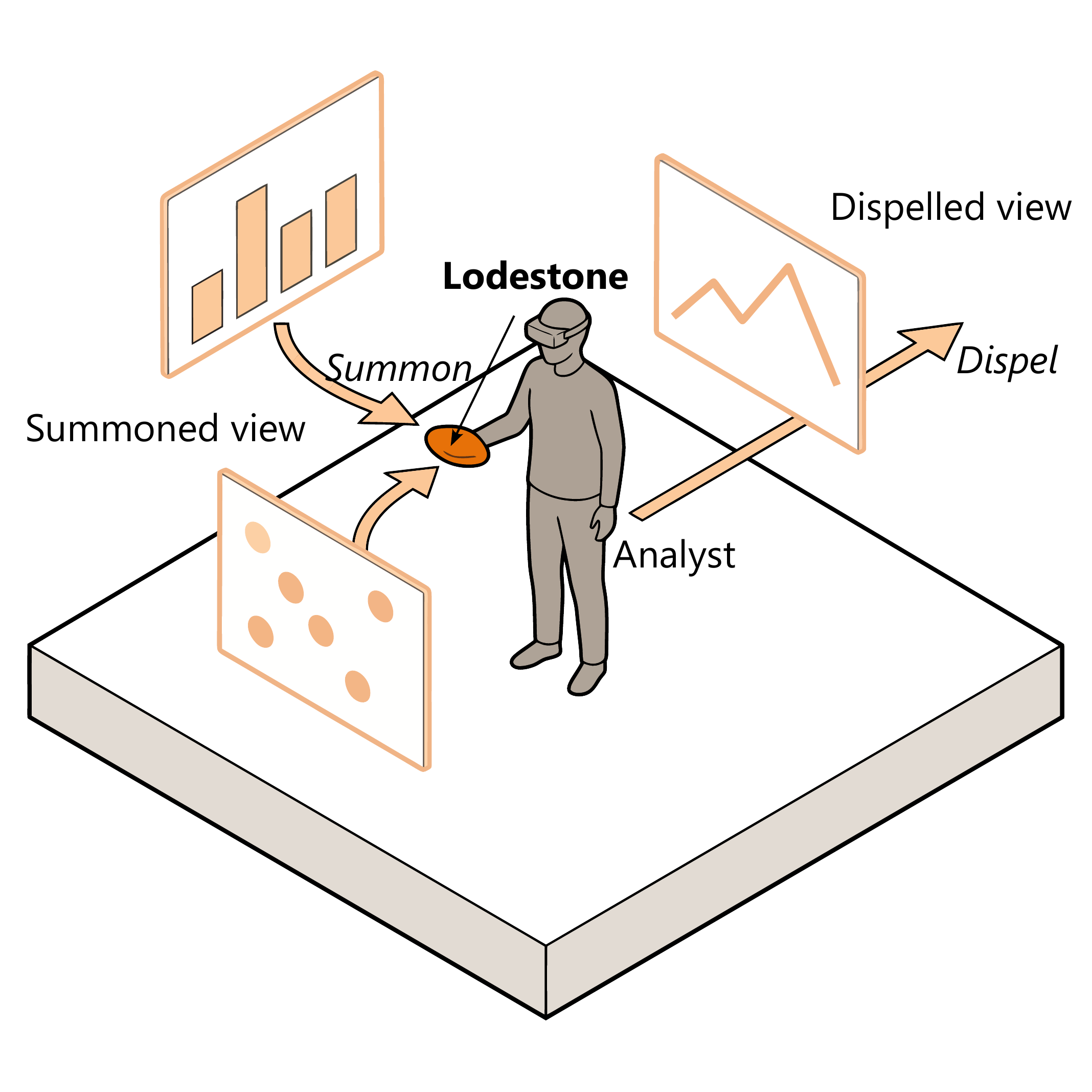}
    \caption{\pat{Lodestone}.
    Holding a lodestone, the analyst draws two chart panels inward and pushes a third away.}
    \label{fig:pat:lodestone}
  \end{subfigure}
  \caption{\textbf{Reach patterns.}
    Each pattern closes the distance between the analyst and the content.
    The analyst acts on distant content through a nearby stand-in~(a), follows a camera path that covers all of it~(b), or draws it in and pushes it away~(c).}
  \label{fig:patterns-reach}
\end{figure*}

Reach patterns close the distance between analyst and content.
The analyst can act on content from afar through a \pat{Marionette}, travel to it along a \pat{Grand Tour}, or bring it closer with a \pat{Lodestone}.

% =====================================================================
% Pattern: Marionette
% Category: Reach
% Source: notes/patterns.md
% =====================================================================

\begin{pattern}{Marionette}{marionette}

\patternsummary{A nearby stand-in that acts on distant or grouped objects.}

\patproblem
The analyst often needs to act on objects that are out of reach, or on a whole group at once, where direct manipulation of each individual object is slow, impractical, or impossible.

\patsolution
Provide a nearby stand-in for the target or the group; manipulating the stand-in manipulates the targets.
The stand-in is also a visual summary of the targets, similar to an interactive legend.

\patconsequences
The coupling has to be legible: the analyst needs to see that acting on one handle acts on everything it stands for, and making that visible remains an open problem in multiple-object interaction.
Marionette adds control to the representational stand-in that the Proxy pattern of Lee et al.~\cite{Lee2024} describes.

\patknownuses
Voodoo Dolls~\cite{DBLP:conf/si3d/PierceSP99} and Surrogate Interaction~\cite{Kwon2011b} both give the user a nearby stand-in that controls a distant or grouped target.
The World in Miniature~\cite{stoakley95virtual} is a Marionette when the analyst uses it to manipulate objects rather than for overview.

\end{pattern}

% =====================================================================
% Pattern: Grand Tour
% Category: Reach
% Source: notes/patterns.md
% =====================================================================

\begin{pattern}{Grand Tour}{grand-tour}

\patternsummary{The system moves the camera along a path that shows everything worth seeing, leaving the scene untouched.}

\patproblem
In immersive data spaces, an analyst often enters a space for the first time or takes it over from another analyst.
This analyst does not yet know the extent of the space or what it contains, and free navigation can leave whole regions unseen.

\patsolution
Compute a camera path that makes every relevant element visible from at least one viewpoint.
Move the analyst along the path, possibly also their view, and keep the scene in place.

\patconsequences
Guided motion trades agency for coverage and can cause discomfort in VR.
Attention data from \pat{Bloom}~(\cref{pat:bloom}) can restrict the tour to what the analyst has not yet seen.
Asimov's grand tour of projections of high-dimensional data~\cite{Asimov1985} is a conceptual ancestor of the pattern and shares its name.

\patknownuses
Elmqvist et al.\ constrain camera motion to support wayfinding in 3D~\cite{DBLP:conf/chi/ElmqvistTT08}, and a short paper at VRST generates tours of 3D environments~\cite{DBLP:conf/vrst/ElmqvistTT07}.
The Batch et al.\ include a Data Tour~\cite{BatchCGF2023}.
Video games such as \textit{World of Warcraft} (2004) use zone flyovers.

\end{pattern}

% =====================================================================
% Pattern: Lodestone
% Category: Reach
% Source: notes/patterns.md
% =====================================================================

\begin{pattern}{Lodestone}{lodestone}

\patternsummary{The analyst attracts content to the body or repels it away, moving the scene instead of the viewer.}

\patproblem
In room-scale AR and VR, and in physical spaces too large to traverse during analysis, content the analyst needs is scattered through the space, and walking to each item is slow.
At the same time, content the analyst does not need crowds the workspace.

\patsolution
Let the analyst summon selected content by attracting it to the body and dispel unwanted content by repelling it.

\patconsequences
Moved content loses its spatial relation to its referent, so the pattern needs a way back, such as a tether or a return gesture.
Lodestone pairs with \pat{Grand Tour}~(\cref{pat:grand-tour}), which moves the viewer to the content along a path.
Go-Go Interaction~\cite{DBLP:conf/uist/PoupyrevBWI96} takes a third approach and extends the user's virtual arm to reach distant objects.

\patknownuses
Bring~\&~Go brings distant nodes to the user when navigating large graphs~\cite{DBLP:conf/chi/MoscovichCHPF09}, and drag-and-pop pulls distant targets toward the cursor~\cite{Baudisch2003b}.
Summoning and dispelling objects were also designed for the system of Batch et al.~\cite{BatchCGF2023}.

\end{pattern}

\subsection{Embodiment}
\label{sec:patterns:embodiment}

% =====================================================================
% Category figure: Embodiment, subfigures (a) to (c)
% Input from contents/patterns.tex at the start of the Embodiment subsection.
% =====================================================================

\begin{figure*}[t]
  \centering
  \begin{subfigure}[t]{\patternwidth}
    \includegraphics[width=\linewidth]{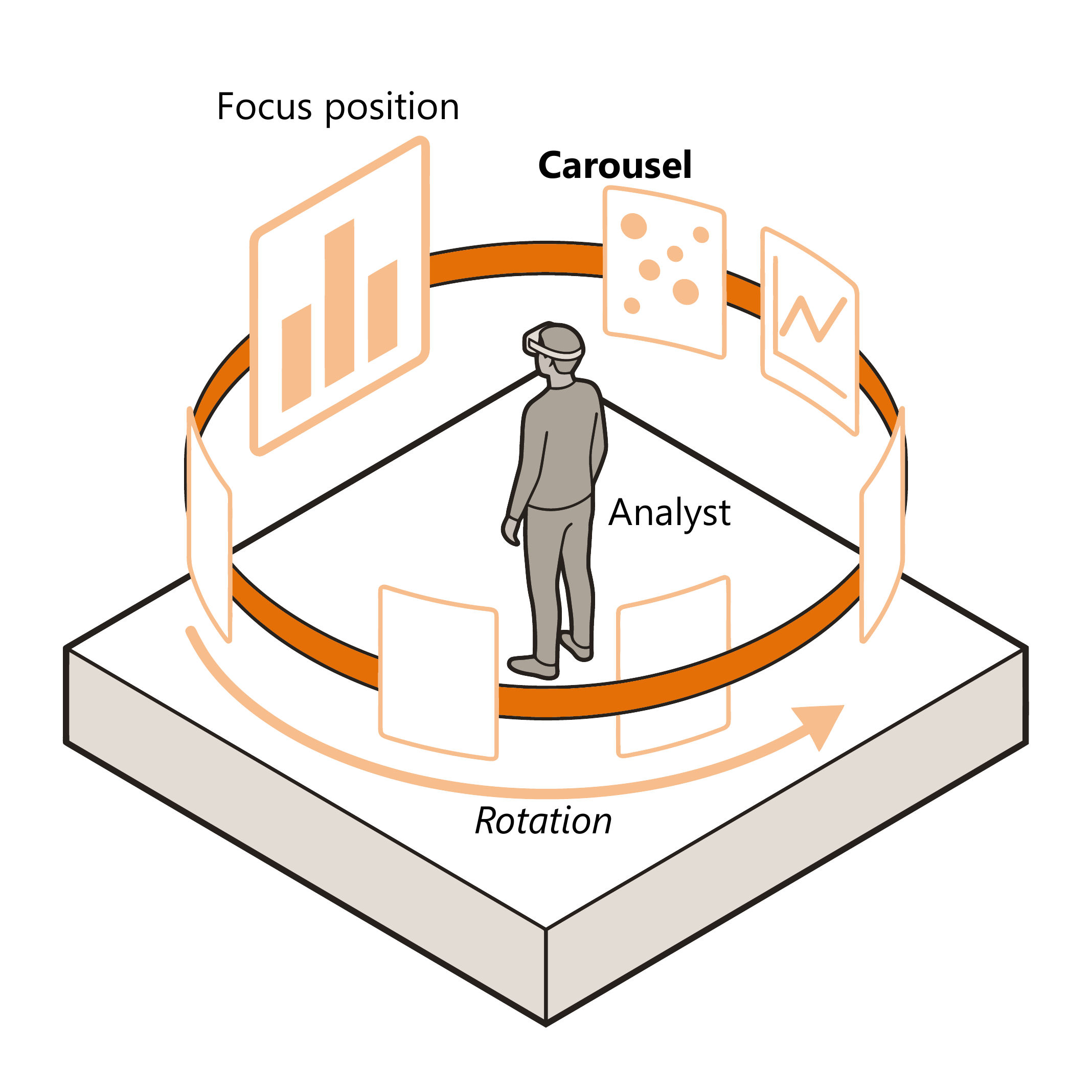}
    \caption{\pat{Carousel}.
    Chart cards ride on a ring around a stationary analyst, and the card in front of the analyst is in focus.}
    \label{fig:pat:carousel}
  \end{subfigure}\hfill
  \begin{subfigure}[t]{\patternwidth}
    \includegraphics[width=\linewidth]{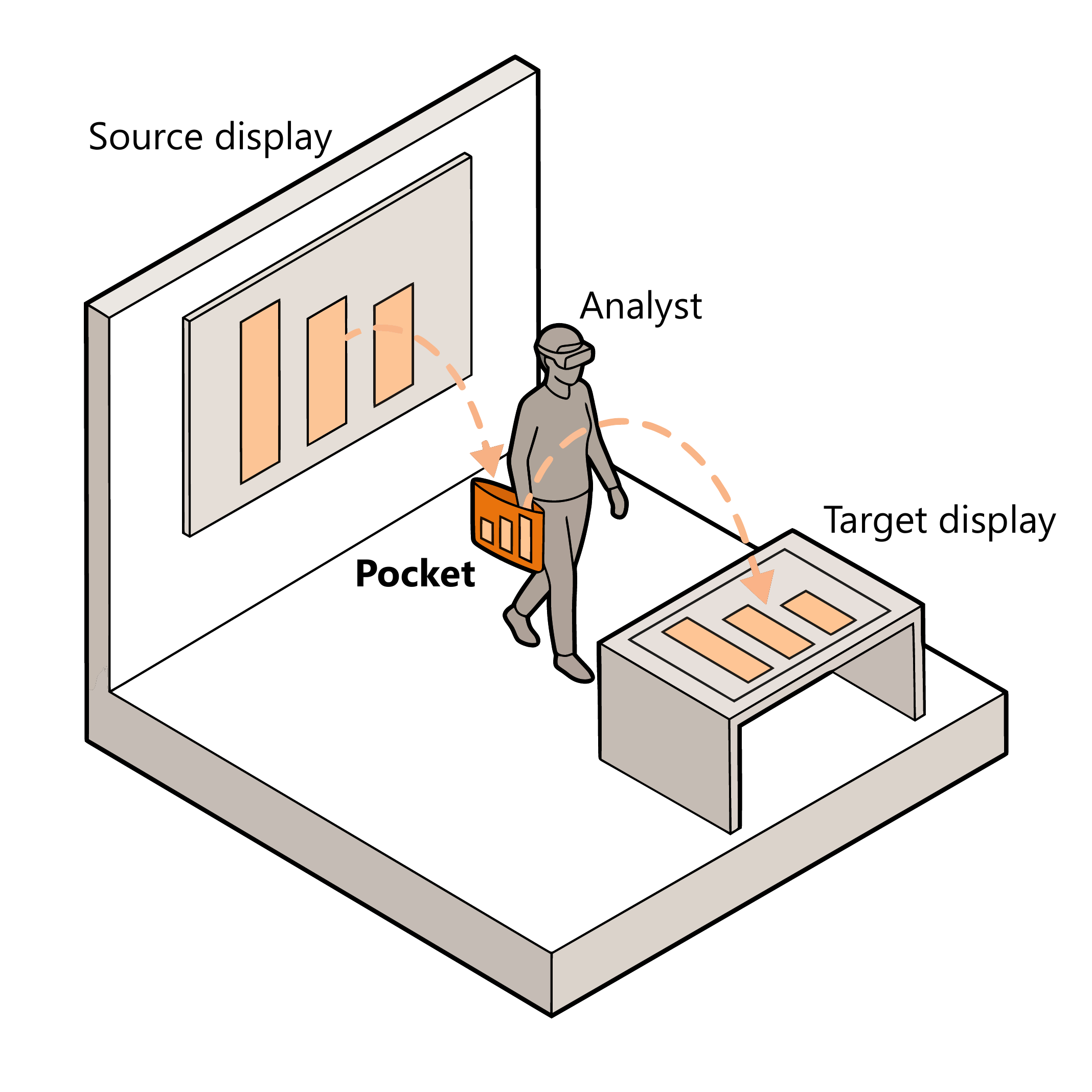}
    \caption{\pat{Pocket}.
    The analyst stores a chart from a wall display in a pocket, walks to a tabletop, and places the chart there.}
    \label{fig:pat:pocket}
  \end{subfigure}\hfill
  \begin{subfigure}[t]{\patternwidth}
    \includegraphics[width=\linewidth]{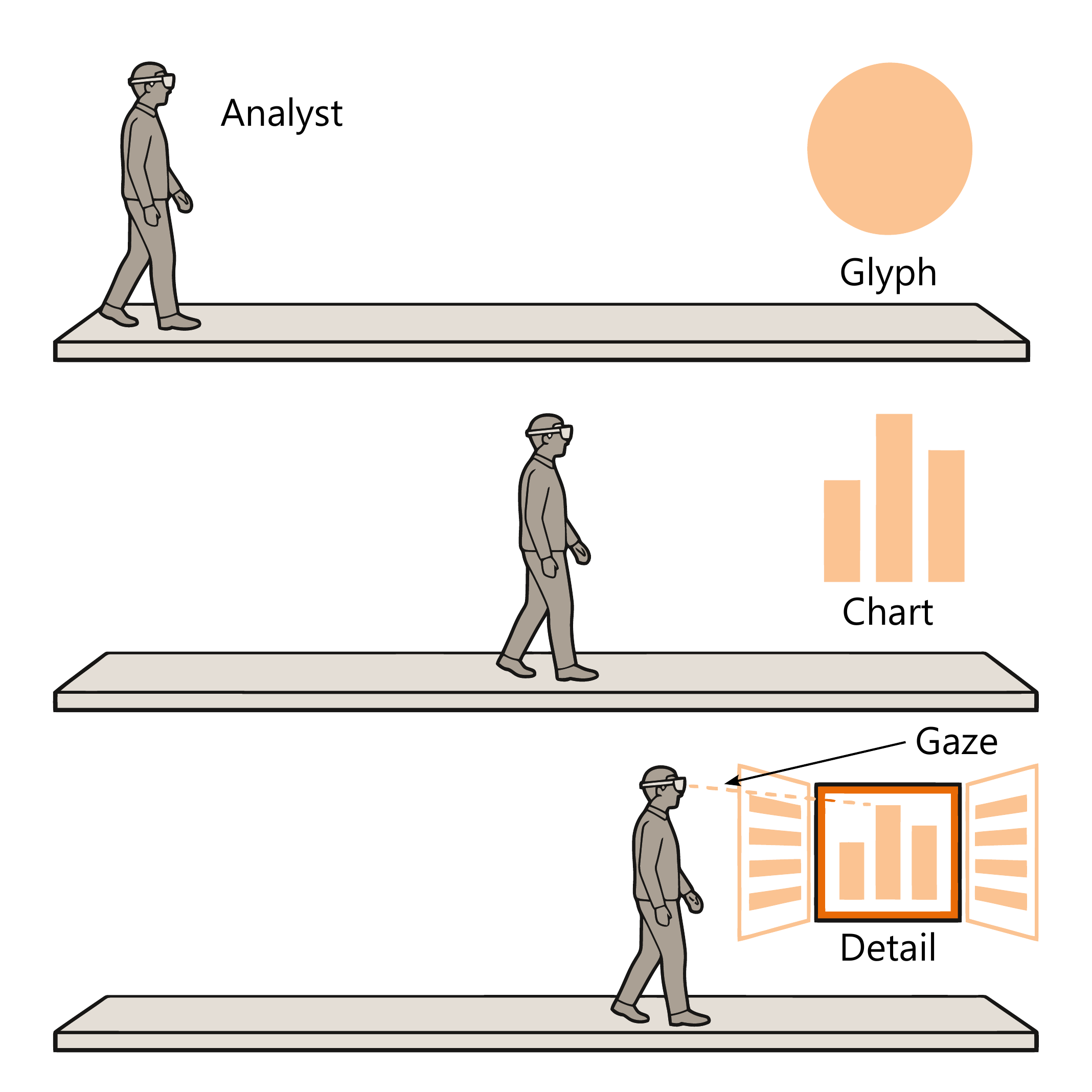}
    \caption{\pat{Bloom}.
    As the analyst walks closer, the same data object grows from a glyph into a chart and then into a detailed view.}
    \label{fig:pat:bloom}
  \end{subfigure}
  \caption{\textbf{Embodiment patterns.}
Each pattern gives the analyst's body a role in the analysis.
The body is the fixed center of a rotating collection~(a), a container that carries views between places~(b), or an implicit input whose approach and gaze open content~(c).}
  \label{fig:patterns-embodiment}
\end{figure*}

Embodiment patterns use the analyst's body as an anchor, a container, or an input.
\pat{Carousel} arranges a collection around a body that stays in place.
\pat{Pocket} turns the body into a container that travels between places and devices, and \pat{Bloom} treats proxemics~\cite{Hall1966}---the position, orientation, and gaze---as implicit input.

% =====================================================================
% Pattern: Carousel
% Category: Embodiment
% Source: notes/patterns.md
% =====================================================================

\begin{pattern}{Carousel}{carousel}

\patternsummary{Items rotate past a fixed viewing position, so the analyst browses a collection without walking.}

\patproblem
Immersive workspaces can hold many charts or cards, and standing or seated analysts have limited room to move.
The analyst must browse a collection of views or items that is larger than the comfortable field of view.
Only part is visible at any moment.

\patsolution
Arrange the items on a ring or cylinder around a fixed viewing position.
Rotate the arrangement to bring items into focus.

\patconsequences
Rotation keeps the analyst still, but the ring hides some items.
The pattern scales poorly beyond a few dozen items.

\patknownuses
Cone Trees arrange the children of each node on a cone that rotates to bring a selected node to the front~\cite{Robertson1991}.
Immersive Economics~\cite{batch20econimmersive} used a Carousel for its 3D axes.
Gallery views of 2D cards in immersive workspaces apply the same layout.

\end{pattern}

% =====================================================================
% Pattern: Pocket
% Category: Embodiment
% Source: notes/patterns.md
% =====================================================================

\begin{pattern}{Pocket}{pocket}

\patternsummary{A personal container carried on the body that holds views and data across places and devices.}

\patproblem
Ubiquitous analytics spans rooms, devices, and sessions, in both mobile and head-worn setups.
Whenever the analyst moves between places or devices, views and data stay behind.

\patsolution
Give the analyst a personal container, anchored to the body, that holds views and data objects.
Content stored in the Pocket on one device or in one place can come out on another.

\patconsequences
The Pocket is private by default, which complicates collaboration; transfer from one person to another needs its own mechanism.
Body-anchored content can also cause clutter.

\patknownuses
David and Goliath uses a smartwatch as a personal store~\cite{Horak2018dng}, and Personal Cockpit anchors views to the body of a head-worn display user~\cite{ens14thepersonalcockpit}.
Pick-and-Drop~\cite{Rekimoto1997} and Stitching~\cite{DBLP:conf/avi/HinckleyRGBS04} move content between devices.
The gesture interface in the film \emph{Minority Report} (2002) is a popular reference for it.

\end{pattern}

% =====================================================================
% Pattern: Bloom
% Category: Embodiment
% Source: notes/patterns.md
% =====================================================================

\begin{pattern}{Bloom}{bloom}

\patternsummary{Content reveals more detail as the analyst approaches or attends to it, and closes again as attention moves on.}

\patproblem
In immersive settings, the system can sense the analyst's position, orientation, and gaze.
The space around the analyst often holds more content than the analyst can read at full detail.
Explicit commands to change the level of detail are disruptive.

\patsolution
Treat the body as implicit input.
Content changes representation with distance, from glyph to chart to table as the analyst approaches, and reveals detail when the analyst attends to it by orientation, gaze, or dwell, closing again as attention moves on.

\patconsequences
A glance does not always signal interest.
The system's inferences should therefore be visible and reversible, which links the pattern to \pat{Trail}~(\cref{pat:trail}).
Attention history can also feed the \pat{Grand Tour} pattern~(\cref{pat:grand-tour}).

\patknownuses
Proxemic interaction adapts content to the distance and orientation of people and devices~\cite{Ballendat2010, Hall1966, Greenberg2011}.
Badam et al.\ apply proxemics to visualization in their proxemic lens~\cite{Badam2016b}.
Pad++ uses semantic zooming, which changes what an object shows as its displayed size changes~\cite{bederson1994}.
Srinivasan et al.\ build attention-aware interfaces~\cite{DBLP:journals/tvcg/SrinivasanEBRE25, Srinivasan2026heedvision}.

\end{pattern}

\subsection{Sensemaking}
\label{sec:patterns:sensemaking}

% =====================================================================
% Category figure: Sensemaking, subfigures (a) to (c)
% Input from contents/patterns.tex at the start of the Sensemaking subsection.
% =====================================================================

\begin{figure*}[t]
  \centering
  \begin{subfigure}[t]{\patternwidth}
    \includegraphics[width=\linewidth]{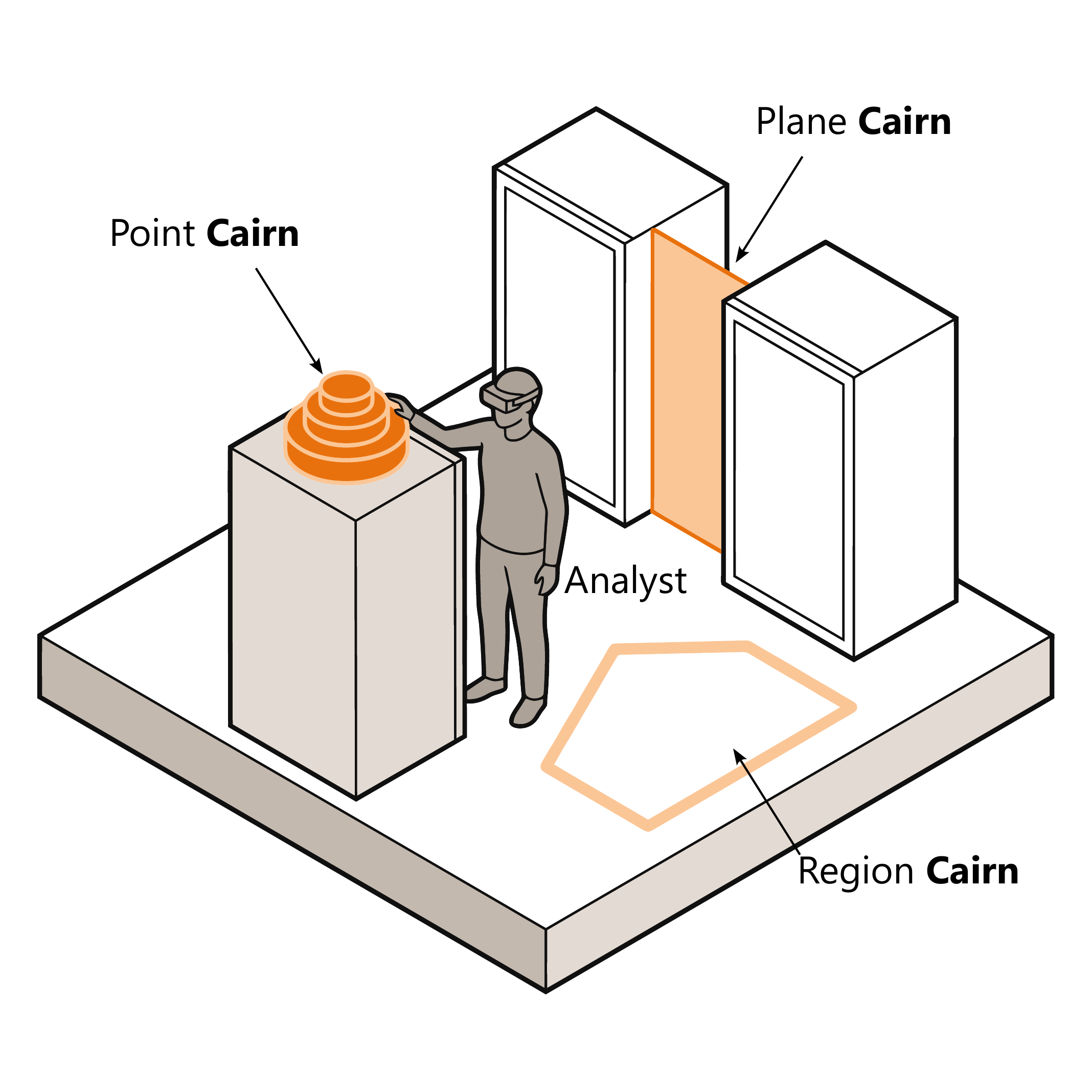}
    \caption{\pat{Cairn}.
An analyst places a point cairn on a machine on a factory floor.
Elsewhere, a region cairn outlines part of the floor and a plane cairn stands between two machines.}
    \label{fig:pat:cairn}
  \end{subfigure}\hfill
  \begin{subfigure}[t]{\patternwidth}
    \includegraphics[width=\linewidth]{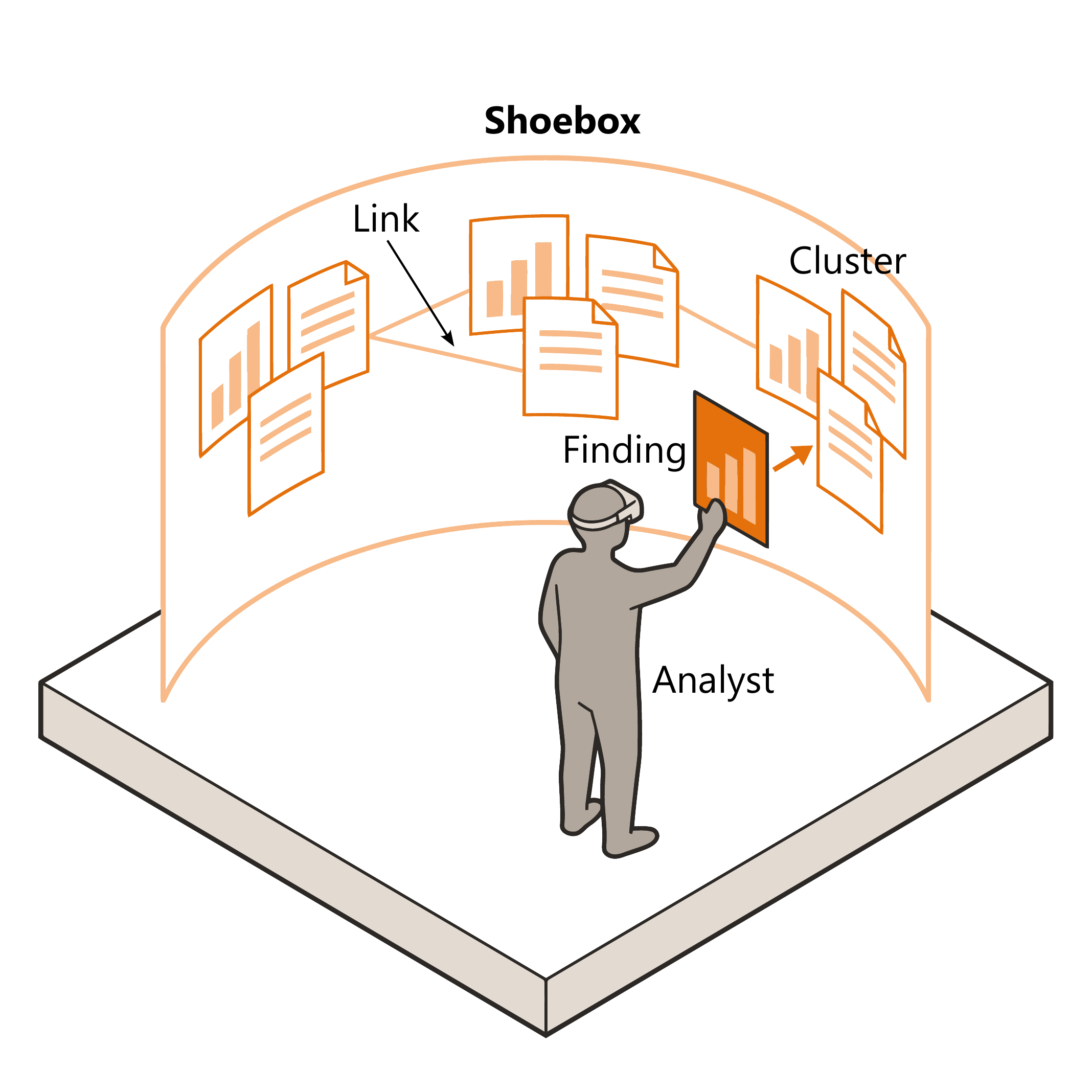}
    \caption{\pat{Shoebox}.
On a large curved surface, the analyst sorts chart and document cards into clusters and links cards across clusters.}
    \label{fig:pat:shoebox}
  \end{subfigure}\hfill
  \begin{subfigure}[t]{\patternwidth}
    \includegraphics[width=\linewidth]{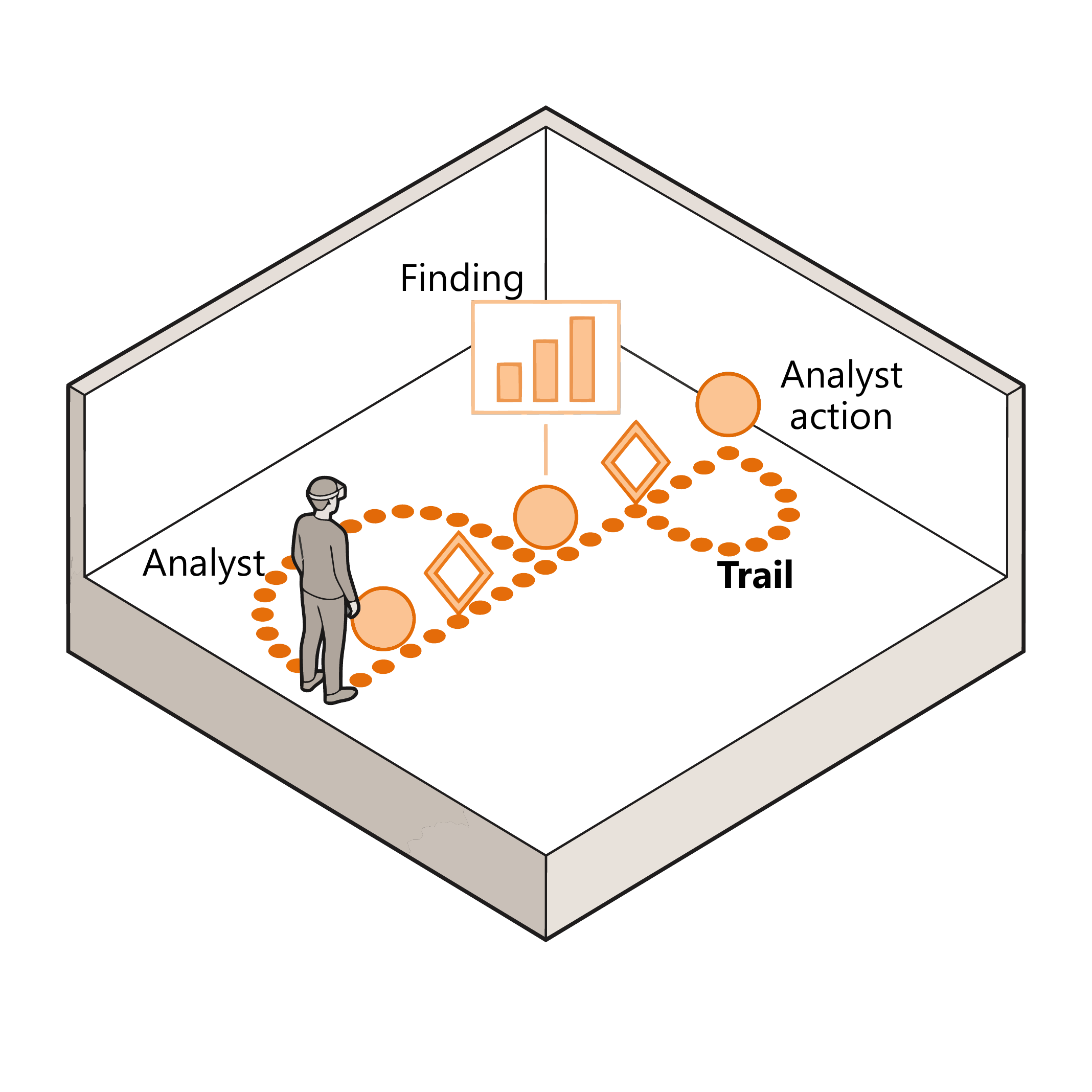}
    \caption{\pat{Trail}.
    A dotted path on the floor records analyst traces.
    Nodes along it mark the analyst's actions and the system's inferences, and one node opens into a snapshot.}
    \label{fig:pat:trail}
  \end{subfigure}
  \caption{\textbf{Sensemaking patterns.}
    Each pattern preserves part of the analysis.
    The analyst marks a finding where it occurs~(a), arranges collected findings into a structure~(b), and retraces the path that led to them~(c).}
  \label{fig:patterns-sensemaking}
\end{figure*}

Sensemaking patterns let the analyst record, organize, and retrace findings.
A \pat{Cairn} records a finding at the place where the analyst made it.
The \pat{Shoebox} gathers findings into a structure, and the \pat{Trail} keeps the sequence of steps that produced them, like Hansel and Gretel's breadcrumbs but out of reach of the birds.

% =====================================================================
% Pattern: Cairn
% Category: Sensemaking
% Source: notes/patterns.md
% =====================================================================

\begin{pattern}{Cairn}{cairn}

\patternsummary{The analyst leaves a persistent mark at a point, region, or plane to record a finding in place.}

\patproblem
In situated and immersive analysis, including geospatial settings, findings are tied to a location in the world or in the data.
These findings are lost when the analyst moves on, since system labels cannot record the analyst's own assertions.

\patsolution
Let the analyst place a persistent mark that records a finding and is anchored to a point, region, or plane.
Like evidence markers at a crime scene, Cairns are numbered or typed so that others can read them later.

\patconsequences
Cairns accumulate and need aggregation or filtering.
Unlike a Label in the sense of Lee et al.~\cite{Lee2024}, which the system attaches to a referent, a Cairn records an assertion that the analyst makes, so placing one is an act of schematization.
Known uses are 2D or sparse, and few of them address analytics in 3D.

\patknownuses
CodeSaw supports spatial messaging~\cite{DBLP:conf/interact/GilbertK07}, and sense.us lets users annotate visualizations graphically~\cite{Heer2007}.
Embodied Provenance brings spatial annotation to IA~\cite{DBLP:journals/pacmhci/ZhangESYG23}.

\end{pattern}

% =====================================================================
% Pattern: Shoebox
% Category: Sensemaking
% Source: notes/patterns.md
% =====================================================================

\begin{pattern}{Shoebox}{shoebox}

\patternsummary{A spatial workspace where the analyst collects findings and arranges them into a structure.}

\patproblem
Some sensemaking tasks involve many documents or findings and take place on large displays or in immersive spaces where layout can carry meaning.
Findings in these tasks stay scattered across views and places.
The analyst needs to collect them and build a structure that explains them.

\patsolution
Provide a spatial workspace where the analyst gathers findings and arranges them into clusters, sequences, or hypotheses.

\patconsequences
Arrangement is itself analysis, so the layout must persist.
Space runs out as findings accumulate, and the structure can diverge from the evidence behind it.

\patknownuses
The shoebox is a stage in the sensemaking loop of Pirolli and Card~\cite{Pirolli2005}.
The Sandbox in nSpace~\cite{Wright2006} and Jigsaw~\cite{Stasko2008} give analysts such a workspace.
Space to Think uses a large display for the same purpose~\cite{Andrews2010}, and its immersive version adapts the approach to VR~\cite{DBLP:conf/vr/LisleDG0B21}.

\end{pattern}

% =====================================================================
% Pattern: Trail
% Category: Sensemaking
% Source: notes/patterns.md
% =====================================================================

\begin{pattern}{Trail}{trail}

\patternsummary{A visible, revisitable record of the analysis path, covering both the analyst's actions and the system's inferences.}

\patproblem
Analyses are often long or interrupted and span places and devices.
Some systems also infer the analyst's intent from body, gaze, or interaction.
As a result, the analyst may lose track of how a finding was reached and of which steps the system took on its own.

\patsolution
Record the analysis path as a visible, revisitable trace in space that includes both the analyst's actions and the system's inferences.
Let the analyst step back along the trace.

\patconsequences
Trails grow long and need summarization.
The designer has to balance the clutter of showing every inference against the loss of trust that hiding them causes.
Trail adds the order of findings in time to the place that a \pat{Cairn}~(\cref{pat:cairn}) marks and the structure that a \pat{Shoebox}~(\cref{pat:shoebox}) builds.

\patknownuses
Bush's Memex links documents through associative trails~\cite{Bush1945}.
VisTrails~\cite{DBLP:conf/visualization/BavoilCSVCSF05} and GraphTrail~\cite{Dunne2012} record the analysis history of visualization users.
StoryFacets also records the analysis path~\cite{DBLP:journals/ivs/ParkSZDRE22}, and Ragan et al.\ characterize provenance in visualization~\cite{DBLP:journals/tvcg/RaganESC16}.
Embodied Provenance applies provenance to immersive analytics~\cite{DBLP:journals/pacmhci/ZhangESYG23}.

\end{pattern}

\section{Designing with the Patterns}
\label{sec:scenario}

Here we demonstrate the use of our pattern catalog by designing a novel tool using them.
Frank, a farm specialist, walks the fields of a client in the morning with smart glasses and a phone, then continues the analysis on an afternoon flight with a mixed-reality headset, a laptop, and a phone.
All day, Frank weighs whether to sow a waterlogged field now or to delay until its drainage is repaired.
In the field, Frank finds a problem whose cause lies underground.
Findings from the field can then later be analyzed offline in an airplane seat.
By evening, Frank needs a case that justifies the decision.
The five mockups in Fig.~\ref{fig:scenario} summarize this scenario.

\subsection{Morning: In the Field}
\label{sec:scenario:morning}

At 07:40 in the morning, Frank stands at the edge of a client's field (Fig.~\ref{fig:scenario}a).
Every field carries a \pat{Label} (a pattern from Lee et al.~\cite{Lee2024}).
The Label of the near field has bloomed into a compact card with soil moisture, soil temperature at 10~cm, and the sowing window; the distant fields stay as glyphs with a name and a dot (\pat{Bloom}, driven by distance).
A \pat{Beacon} at the right edge of the view points to Field~7, which lies 410~m away and out of view, and reports high moisture there.
Dashed lines preview the planned sowing rows (\pat{Ghost}~\cite{Lee2024}).
The phone carries what the glasses cannot show legibly, a 14-day chart of soil moisture.

At 08:30, Frank has followed the Beacon to a wet spot in Field~7 (Fig.~\ref{fig:scenario}b).
The ground opens into a pit about 80~cm deep that exposes the buried drain, with one segment highlighted and labeled with its flow: 0.2~L/s against 1.4~L/s expected.
The data decides what shows through, since the cutaway appears only where the flow sensor reports a shortfall (\pat{X-Ray}).
Frank places a numbered flag, ``3,'' with the note ``Blocked drain? Delay sowing Field 7'' (\pat{Cairn}).
A tray above the wrist holds four items, the newest of them a copy of the flag, so the finding leaves the field with Frank (\pat{Pocket}).

\subsection{Afternoon: In Transit}
\label{sec:scenario:afternoon}

At 14:20, Frank reviews the morning's field work in the airplane on the way home (Fig.~\ref{fig:scenario}c).
Chart panels (\pat{Panel}~\cite{Lee2024}) are arranged in an arc around Frank's head and rotate past, since the seat does not let Frank turn (\pat{Carousel}).
The \pat{Pocket} opens above the wrist with the flag and note from 08:30.
Frank pulls the drainage map of Field~7 out of the ring toward the body (\pat{Lodestone}).
A line from the laptop shows where one of the panels came from.

At 14:50, a miniature of the farm stands on the tray table (\pat{Proxy}~\cite{Lee2024}; Fig.~\ref{fig:scenario}d).
Three orthographic views, a plan and two elevations, unfold around it.
Leader lines tie the faulty drain segment across all views (\pat{Shadow Box}).
A  sphere in Frank's hand pushes data aside to expose the drain (\pat{Sieve}).
Beside the model, three crop tokens each act on every field of their crop, so pinching ``Winter wheat'' selects all wheat fields at once (\pat{Marionette}).

At 15:30, Frank builds the case (Fig.~\ref{fig:scenario}e).
A board above the tray holds cards in three clusters, for drainage, soil temperature, and the sowing window; the morning's photo and flag note sit among them (\pat{Shoebox}).
A timeline along the bottom of the board records the day, with solid dots for Frank's actions and hollow dots for the system's inferences (\pat{Trail}).
One hollow dot reads ``Inferred interest: Field 7'' and offers an undo control.
On the tray, a small model offers a tour of the three fields Frank has not yet visited; because the real farm is out of reach, this \pat{Grand Tour} plays inside a \pat{Proxy}.
The board now sets the evidence on drainage beside the evidence on soil temperature and the sowing window.
From it, Frank can argue for sowing Field~7 now or for waiting until the drain is repaired.

\section{Discussion}
\label{sec:discussion}

\paragraph{Extending a shared language.}

The design scenario in \cref{sec:scenario} combined all thirteen of our patterns with four of those of Lee et al.~\cite{Lee2024} in one design.
It demonstrated that the analyst's setting decides which patterns apply.
Our patterns go past visual representation into reach, embodiment, and sensemaking.
The language also covers mobile and cross-device settings, as well as virtual data spaces with no physical referent.
The thirteen patterns are a starting set, much as the twenty-three of the Gang of Four~\cite{Gamma1994} were followed by decades of further catalogs, from software architecture to interaction design~\cite{Borchers2001} and interface design~\cite{Tidwell2005}.
Others can contribute new patterns to this language, refine the entries we give, and challenge the four categories themselves.

\paragraph{A vocabulary for design.}

Pattern names let designers compose and discuss solutions without restating them.
Software developers say ``traverse the Composite with an Iterator,'' allowing colleagues to understand without an explanation of either.
Some patterns have become infrastructure: Java's \texttt{Iterator} and \texttt{Iterable} interfaces underlie the enhanced \texttt{for} loop, so the pattern is now part of how the language works.
We could see a similar trajectory in IA toolkits.

Names also serve generative use: a designer facing a problem consults the matching category.
In the design scenario, the buried drain led to sight, where a \pat{Beacon} pointed Frank to Field~7 and the Consequences of \pat{X-Ray} and \pat{Sieve} settled which of the two belonged in the field and which in the seat.

\paragraph{A domain-specific pattern language.}

A traditional visualization designer tends to have knowledge in tools such as D3~\cite{Bostock2011}, Vega-Lite~\cite{Satyanarayan2017}, or Tableau.
When building for a head-worn display, they will encounter coordinate frames, transformation matrices, quaternions, scene graphs, shaders, and full game engines~\cite{DBLP:journals/cga/ButcherBSMER24}.
Most of that machinery is beside the point for building a visualization; much of it is a red herring.
Domain-specific languages raise the level of abstraction for one domain.
In this paper, we offer a domain-specific pattern language that names solutions above any particular engine or toolkit, so designers can build these tools without mastering every discipline involved.

\paragraph{Limitations.}

Patterns are distilled experience: Alexander wrote from architectural practice~\cite{Alexander1979, Alexander1977}, while Beck and Cunningham~\cite{Beck1987} and the Gang of Four~\cite{Gamma1994} were working software engineers.
Expert curation is therefore the original method of building a pattern language, regardless of topic.
It trades the coverage of a systematic review for depth of interpretation.

Literature mining came later and complements curation: Lee et al.\ coded 457 papers~\cite{Lee2024}, and Elmqvist and Tsigas surveyed occlusion techniques for their taxonomy~\cite{Elmqvist2008}.
Taken to its extreme would have been to systematically identify interaction techniques in a large corpus of papers and then mechanically cluster them using a k-means or UMAP cluster approach similar to the ensemble clustering done in the situated analytics survey by Shin et al.~\cite{Shin2024}.
This would potentially have allowed us to automatically identify the most common archetypes in the literature. 
However, a design pattern is a fundamentally empirical construct that draws on its author's prior knowledge and experience.

\section{Conclusion}
\label{sec:conclusion}

Designers who build visualizations for glasses, headsets, phones, and the spaces between them need solutions they can name, compare, and combine.
We proposed thirteen design patterns for ubiquitous analytics in four categories (sight, reach, embodiment, and sensemaking), which extend the situated-visualization patterns of Lee et al.~\cite{Lee2024} from how data is represented to how the analyst sees, reaches, uses the body, and makes sense of findings.
A design scenario that follows one farm manager from the field to an airplane seat showed the two languages composing within a single design, with the setting deciding which patterns apply.
The language can grow as new recurring patterns are identified in future applications.

\section*{Disclosure of Generative AI Content}
\label{sec:genai-disclosure}

We generated the five scenario mockups in \cref{fig:scenario} with Google Gemini 3.8 Flash Image to illustrate the design in \cref{sec:scenario}.
The other images were generated using Gemini, and later edited and annotated by the authors.
Generative AI (Anthropic Claude Opus 5.5) was used to generate some sections, including text, tables, and formatting. 
The authors take full responsibility for the content.

\acknowledgments{%
    This work was supported by Villum Investigator grant VL-54492 by Villum Fonden.
    Any opinions, findings, and conclusions expressed in this material are those of the authors and do not necessarily reflect the views of the funding agency.
}

\bibliographystyle{abbrv-doi}
\bibliography{aaa-patterns}

%% arXiv only: the supplemental working list as an appendix. The same body
%% file is input by supplemental.tex, so the two versions cannot drift apart.
\ifarxiv
  \appendix
  \section{Working List of Pattern Candidates}
  \label{sec:appendix:list}
  % --------------------------------------------------------------------------
% supp-working-list.tex -- body of the supplemental working list.
%
% Input by supplemental.tex (standalone supplement, under \section*{Working List})
% and by aaa-patterns.tex in arXiv mode (Appendix A). No heading here.
% Citations are author-year text, since the standalone supplement has no
% reference list.
% --------------------------------------------------------------------------

We curated the thirteen patterns from a working list of 39 entries: the ten patterns of Lee, Sedlmair, and Schmalstieg (2024), which served as a baseline, and 29 candidates of our own.
Our candidates came from our own systems, including four occlusion techniques from Elmqvist's dissertation (2008) and five ideas from the VR system of Batch et al.\ (2023), and from the visualization, AR, and HCI literature.
Each candidate had to clear the closest prior pattern, and Table~\ref{tab:lee} records how Lee et al.'s patterns relate to ours.

\begin{table}[htb]
  \caption{\textbf{Existing patterns.} Lee et al.'s ten patterns and their relation to ours.}
  \label{tab:lee}
  \centering
  \footnotesize
  \begin{tabularx}{\columnwidth}{@{}l>{\raggedright\arraybackslash}p{0.30\columnwidth}>{\raggedright\arraybackslash}X@{}}
    \toprule
    \textbf{Pattern} & \textbf{Summary} & \textbf{Relation to our candidates and patterns} \\
    \midrule
    \pat{Glyph}          & Data mark touching the referent & Covers Referent Overlay \\
    \pat{Decal}          & Dense texture on a physical surface & Covers Referent Overlay \\
    \pat{Label}          & Added information attached to a referent & \pat{Cairn} differs in authorship \\
    \pat{Trajectory}     & Line linking points or showing motion & Covers Tether \\
    \pat{Ghost}          & Virtual object resembling a physical one & \\
    \pat{Panel}          & Conventional chart semantically linked to a referent & Covers Satellite \\
    \pat{Magic Lens}     & Region that alters what it covers & The \pat{Sieve} clears what it covers; restyling belongs to \pat{Magic Lens} \\
    \pat{Morph}          & Physical objects reshaped to reduce clutter & Neighbor of the \pat{Sieve}'s displacement variant \\
    \pat{Proxy}          & Nearby replica of a distant referent & Covers Minimap; \pat{Marionette} adds control; \pat{Shadow Box} adds unfolding \\
    \pat{Virtual Mirror} & Reflected view with a data overlay & None \\
    \bottomrule
  \end{tabularx}
\end{table}

Table~\ref{tab:candidates} lists our 29 candidates in eight groupings with their fate.
Kept means the candidate became the root of a final pattern, possibly under a new name; Absorbed means it merged into a survivor; Deferred means prior work already covers it, and we credit that work; Dropped means we removed it; Open means it was not decided before the set closed.

\begin{table*}[tb]
  \caption{\textbf{Candidate patterns.} Our 29 working-list candidates and their fate.}
  \label{tab:candidates}
  \centering
  \footnotesize
  \begin{tabularx}{\textwidth}{@{}r>{\raggedright\arraybackslash}p{0.17\textwidth}l>{\raggedright\arraybackslash}Xll@{}}
    \toprule
    \textbf{\#} & \textbf{Candidate} & \textbf{Grouping} & \textbf{Description} & \textbf{Fate} & \textbf{Outcome} \\
    \midrule
    1  & Referent Overlay & Reference & Data registered directly onto the referent's surface & Deferred & Lee et al.'s \pat{Glyph} and \pat{Decal} \\
    2  & Satellite & & Chart floating beside its referent & Deferred & Lee et al.'s \pat{Panel} \\
    3  & Tether & & Connector, such as a leader line, binding a mark to its referent & Deferred & Lee et al.'s \pat{Trajectory} \\
    4  & \pat{Beacon} & & In-view marker pointing to a referent out of view & Kept & \pat{Beacon} (Sight) \\
    5  & Situated Representation & & Representation near its referent but detached from it & Deferred & Willett et al.\ (2017) \\
    6  & Embedded Representation & & Representation that is part of its referent & Deferred & Willett et al.\ (2017) \\
    \midrule
    7  & Approach Gradient & Representation & Encoding switches with viewing distance, as in semantic zooming & Absorbed & \pat{Bloom} \\
    8  & Billboard & & Representation rotates to face the viewer; a toolkit default & Dropped & \\
    9  & Thinning & & Only what fits is shown; the rest is suppressed & Absorbed & \pat{Sieve} (selection slot) \\
    \midrule
    10 & Projection Distorter & Occlusion & Perspective and orthographic projection blended to reveal hidden objects (Elmqvist, 2008) & Dropped & \\
    11 & Balloon Probe & & Probe that pushes objects apart (Elmqvist, 2008); generalized to any moving volume & Kept & \pat{Sieve} (Sight) \\
    12 & Dynamic Transparency & & Occluders turn transparent on demand (Elmqvist, 2008); in AR, the data decides & Kept & \pat{X-Ray} (Sight) \\
    13 & Off-Screen and Occluded-Object Labeling & & Signals for hidden or out-of-view objects (Elmqvist, 2008) & Absorbed & \pat{Beacon} \\
    \midrule
    14 & World in Miniature & Scale and Overview & Small 3D copy of the space; its 2D Minimap variant went to Lee et al.'s \pat{Proxy} & Kept & \pat{Marionette} (Reach) \\
    15 & \pat{Shadow Box} & & 3D scene unwrapped into flat views (Batch et al., 2023) & Absorbed & \pat{Shadow Box}, via Projection \\
    16 & Projection & & Scene unfolded into orthographic views around it & Kept & \pat{Shadow Box} (Sight) \\
    \midrule
    17 & Proxemic Control & Inference & Body position drives the view; includes its Gaze Preview sub-item & Kept & \pat{Bloom} (Embodiment) \\
    18 & Inference Receipt & & Visible, reversible trace of each system inference & Kept & \pat{Trail} (Sensemaking) \\
    \midrule
    19 & Frame Lock & Body and Space & Each element's reference frame as a switchable property & Open & \\
    20 & Lazy Susan & & Items on a ring rotate past a fixed viewer & Absorbed & \pat{Carousel} (Embodiment) \\
    21 & Surface Claim & & Content placed by allocating surfaces in the environment & Open & \\
    22 & Handoff & & Analytical state transferred between devices & Absorbed & \pat{Pocket} \\
    23 & Heads-Up Display & & Content fixed to the camera & Open & \\
    24 & Body-Carried Container & & Personal store for views and data, carried on the body & Kept & \pat{Pocket} (Embodiment) \\
    25 & Summoning and Dispelling & & Content attracted to or repelled from the body (designed for Batch et al., 2023) & Kept & \pat{Lodestone} (Reach) \\
    26 & Data Tour & & Camera path that shows every relevant element (Batch et al., 2023) & Kept & \pat{Grand Tour} (Reach) \\
    \midrule
    27 & Space versus Place & Place & Distinction between a location and a location with meaning & Open & \\
    \midrule
    28 & \pat{Cairn} & Sensemaking & Analyst's persistent mark recording a finding in place & Kept & \pat{Cairn} (Sensemaking) \\
    29 & \pat{Shoebox} & & Spatial workspace for collecting and arranging findings & Kept & \pat{Shoebox} (Sensemaking) \\
    \bottomrule
  \end{tabularx}

  \par\smallskip\raggedright\footnotesize
  Table note: 13 kept, 5 absorbed, 5 deferred, 2 dropped, and 4 open.
  The list's two sub-items, Minimap and Gaze Preview, are counted with their parent entries.
\end{table*}

\fi

\end{document}